\documentclass[]{tJOT2e}

\usepackage{graphicx} 
\usepackage{dcolumn}
\usepackage{bm}
\usepackage{amssymb}
\usepackage{amsmath}
\usepackage{wasysym}
\usepackage{color}
\usepackage{hyperref}
\usepackage{float}
\usepackage{flushend}
\usepackage{soul}
\usepackage{hyperref}
\hypersetup{
    colorlinks,%
    citecolor=blue,%
    filecolor=blue,%
    linkcolor=blue,%
    urlcolor=blue
} 

\begin{document}

\markboth{R. Kumar, M. K. Verma, and R. Samtaney}{Journal of Turbulence}

\articletype{}

\title{Energy transfers in dynamos with small magnetic Prandtl numbers}
\vskip 0.3 in

\author{Rohit Kumar$^{\rm a \ast}$, \thanks{$^\ast$Corresponding author. Email: rohitkr@iitk.ac.in
\vspace{6pt}} Mahendra K. Verma$^{\rm a}$, and 
\vspace{6pt} Ravi Samtaney$^{\rm b}$\\\vspace{6pt}  $^{\rm a}${\em{Department of Physics, Indian Institute of Technology, Kanpur 208016, India}}\\$^{\rm b}${\em{Mechanical Engineering, Division of Physical Sciences and Engineering, King Abdullah University of Science and Technology Thuwal 23955-6900, Kingdom of Saudi Arabia}} }


\maketitle

\begin{abstract} 
We perform numerical simulation of dynamo with magnetic Prandtl number $\mathrm{Pm} =0.2$ on $1024^3$ grid, and compute  the energy fluxes and the shell-to-shell energy transfers. These computations indicate that the magnetic energy growth takes place mainly due to the energy transfers from large-scale velocity field to large-scale magnetic field and that the magnetic energy flux is forward.  The steady-state magnetic energy is much smaller than the kinetic energy, rather than equipartition; this is because the magnetic Reynolds number is near the dynamo transition regime.  We also contrast our results with those for dynamo with $\mathrm{Pm} =20$ and decaying dynamo.
\end{abstract}


\section{Introduction}
\label{sec:intro}

Self-generation of a magnetic field in a conducting fluid is called  {\em dynamo effect}~\cite{Moffatt:book};  such phenomena is observed in astrophysical objects including planets, stars, and galaxies.  Some of the important parameters for the dynamo action are the magnetic Prandtl number $\mathrm{Pm} = \nu/\eta$, and magnetic Reynolds number $\mathrm{Rm}=UL/\eta$, where $U$ and $L$ are the characteristic velocity and length, respectively,  of the system, and $\nu$ and $\eta$ are the kinematic viscosity and magnetic diffusivity, respectively, of the fluid.   The magnetic Prandtl number in natural dynamos are either very low or very high. For example, in Earth's core and in liquid sodium experiments, $\mathrm{Pm} \sim 10^{-6}$, whereas in galaxies, $\mathrm{Pm} \sim 10^{6}$. 
  
 The magnetic Prandtl number plays a major role in the dynamo process.  Theoretical arguments, numerical simulations, and experiments  reveal that for $\mathrm{Pm} >1$ (e.g., Galactic plasma), the magnetic field typically grows at characteristic length scales smaller than that of the velocity field; this process is known as the small-scale dynamo (SSD)~\cite{Brandenburg:PR2005,Cattaneo:JFM2002}. On the other hand, in systems for which $\mathrm{Pm} <1$ (e.g., liquid metal, Earth's core), the magnetic field generally grows at characteristic length scales of the order of or larger than that of the velocity field~\cite{Moffatt:book,Brandenburg:PR2005,Monchaux:PRL2007}.    When the length scale of the magnetic energy growth  is larger than that of the kinetic energy, the process is called the large-scale dynamo (LSD).  Typically, kinetic and magnetic helicities induce LSD.  Solar dynamo, which has $\mathrm{Pm} <1$ is an example exhibiting both SSD and LSD~\cite{Cattaneo:JFM2002}.  


\begin{figure}[htbp]
\centering
\includegraphics[scale=0.5]{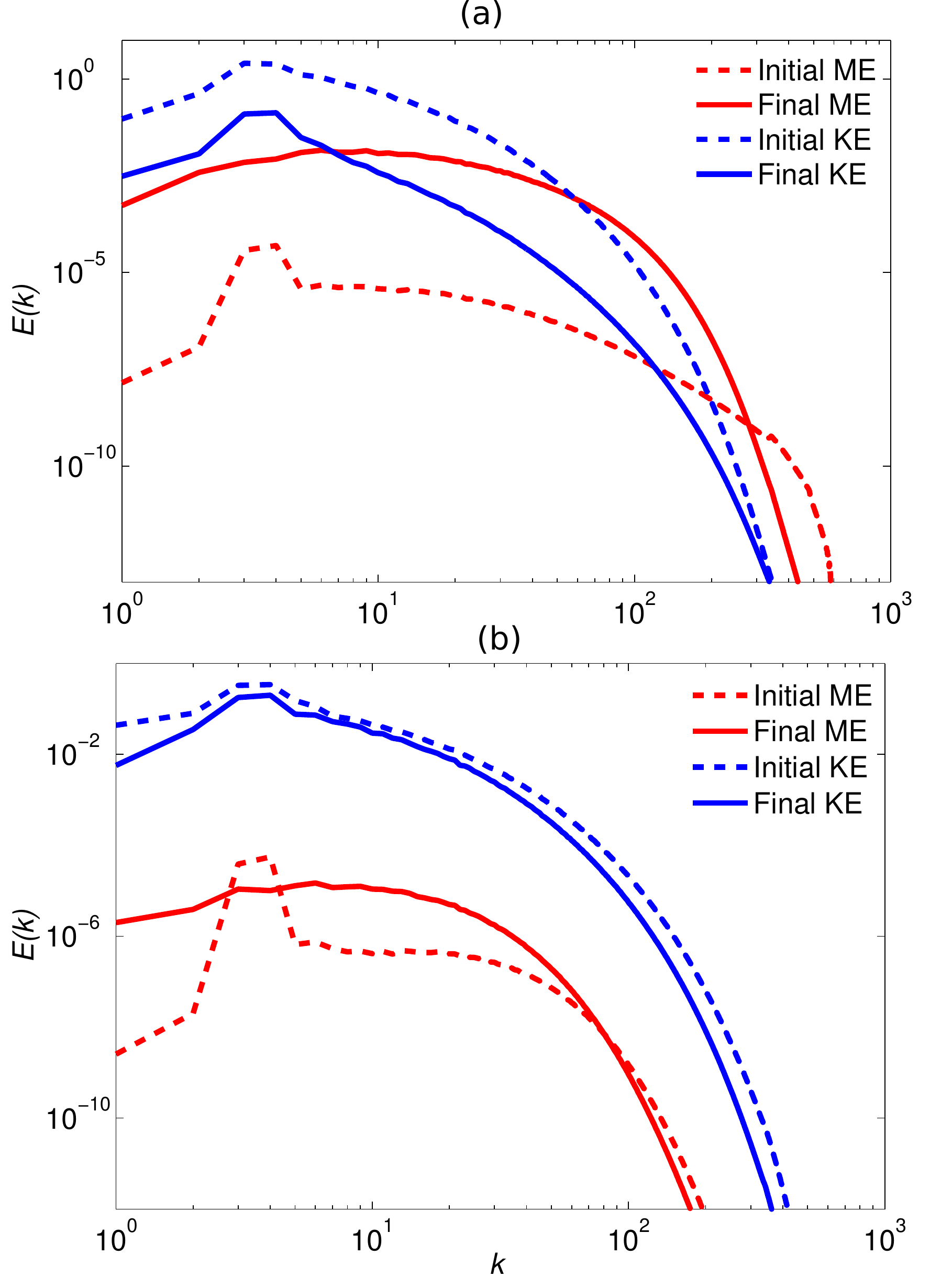} 
\caption{Plots of kinetic (KE, blue curves) and magnetic (ME, red curves) energy spectra for (a) dynamo with $\mathrm{Pm} =20$ ($\nu =0.01$ and $\eta =0.0005$), and (b) dynamo with $\mathrm{Pm} =0.2$ ($\nu =0.002$ and $\eta =0.01$).  The initial spectra, shown as dashed curves, evolve to the final spectra exhibited by solid curves.}
\label{fig:Ek_ssd_lsd}
\end{figure}


\begin{figure}[htbp]
\centering
\includegraphics[scale=0.35]{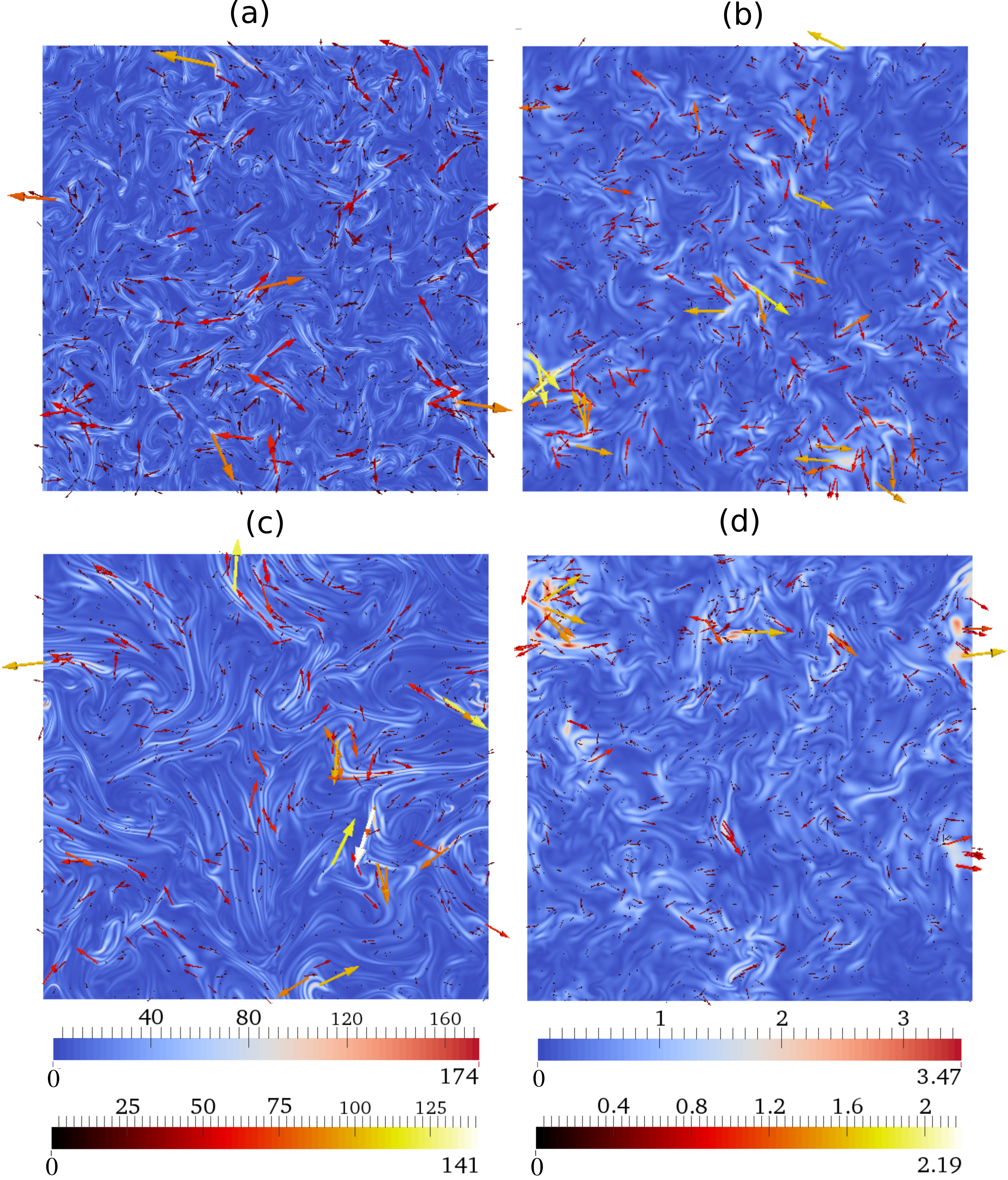} 
\caption{Density plots of the current density ${\mathbf J} = \nabla \times {\mathbf b}$ of a cross section: for dynamo with $\mathrm{Pm} =20$ (a) at $t=1.15$, (c) at $t=18.73$; and $\mathrm{Pm} =0.2$, (b) at $t=2.29$, (d) at $t=16.24$. The vector ${\mathbf J}_\perp = (J_x,J_y)$ are superposed on the density plots.   The first colour bar is for the  density plots, while the latter for the vector plots.  Intense currents ${\mathbf J}_\perp$ are shown using red and yellow vectors.  In the early stages of dynamo for $\mathrm{Pm} =20$, strong currents are present at small scales, while for $\mathrm{Pm} =0.2$ relatively large-scale currents are observed.}
\label{fig:real_ssd_lsd}
\end{figure}

A typical dynamo process is studied by starting a magnetohydrodynamic (MHD) flow  with a small seed magnetic field, and then the growth and saturation of the magnetic field is investigated.   In Fig.~\ref{fig:Ek_ssd_lsd}(a,b), we illustrate the evolution of the kinetic and magnetic energy spectra of typical nonhelical (vanishing kinetic and magnetic helicities) dynamos with $\mathrm{Pm} >1$ and $\mathrm{Pm} <1$ respectively.  Figure~\ref{fig:Ek_ssd_lsd}(a) depicts how magnetic energy (the red curves) grows at small length scales (or large wavenumbers)  for $\mathrm{Pm} =20$.  Figure~~\ref{fig:Ek_ssd_lsd}(b) depicts the magnetic energy growth at large length scales for $\mathrm{Pm} =0.2$.  These features are also evident in real-space depiction of the magnitude of the current density ${\mathbf J} = \nabla \times {\mathbf b}$ in Fig.~\ref{fig:real_ssd_lsd}; here ${\mathbf b}$ is the magnetic field. The initial current density for $\mathrm{Pm} =20$ (Fig.~\ref{fig:real_ssd_lsd}(a)) shows small islands of intense  $|{\mathbf J}|$, while the plot for $\mathrm{Pm} =0.2$  (Fig.~\ref{fig:real_ssd_lsd}(b)) shows relatively large islands of intense $|{\mathbf J}|$.  The final configurations of ${\mathbf J}$ shown in Fig.~\ref{fig:real_ssd_lsd}(c) and Fig.~\ref{fig:real_ssd_lsd}(d) for $\mathrm{Pm} =20$ and $0.2$, respectively, exhibit large-scale currents. 

For the systems with $\mathrm{Pm} >1$ or $\nu > \eta$,  the length scales of magnetic energy dissipation is smaller than  that for the kinetic energy (See Schekochihin {\it et al.}~\cite{Schekochihin:APJ2004}). For $\mathrm{Pm} <1$ or $\nu < \eta$, the inequality is reversed.    When both the velocity and magnetic fields are turbulent, the viscous length scale ($l_{\nu}$) and resistive length scale ($l_{\eta}$) are defined as $l_{\nu} = \left(\nu^3/ \epsilon_i\right)^{1/4}$ and $l_{\eta} = \left(\eta^3/ \epsilon_i/\right)^{1/4}$, where $\epsilon_i$ is the kinetic energy supply rate, hence $l_\nu/l_\eta \sim \mathrm{Pm}^{3/4}$.  For $ \mathrm{Pm} > 1$, we obtain $l_{\nu} > l_{\eta}$ and vice versa. Note however that for the figures~\ref{fig:Ek_ssd_lsd}(a) and~\ref{fig:Ek_ssd_lsd}(b), the kinetic and magnetic fields are not turbulent.

Dynamo process has been studied using numerical simulations.  Chou~\cite{Chou:APJ2001b} and Schekochihin {\it et al.}~\cite{Schekochihin:APJ2004,Schekochihin:PRL2004b} simulated dynamo for  $\mathrm{Pm} > 1 $ and observed exponential growth of magnetic energy in time, which is dominated at small scales. Ponty {\it et al.}~\cite{Ponty:PRL2005} studied dynamo transition for $\mathrm{Pm} <1$ using Taylor-Green forcing and observed growth of the kinetic and magnetic energy spectra until saturation, during which the magnetic energy attains an approximate equipartition with the kinetic energy. Mininni and Montgomery~\cite{Mininni:PRE2005b} presented results of dynamo transition at low magnetic Prandtl numbers using Roberts flow (helical) and obtained similar results. Ponty and Plunian~\cite{Ponty:PRl2011} studied transition between small-scale and large-scale dynamo by varying the Prandtl number.  Stepanov and Plunian~\cite{Stepanov:JT2006,Stepanov:APJ2008,Plunian:PR2013} studied the dynamo action using shell models and obtained interesting results; shell models enabled them to investigate dynamo action for very small and very large Prandtl numbers.   

The energy transfer processes, e.g., the energy fluxes and shell-to-shell energy transfers, provide valuable information on the dynamo process, which has been well studied for $\mathrm{Pm}=1$~\cite{Dar:PD2001,Verma:PR2004,Debliquy:PP2005,Carati:JT2006,Alexakis:PRE2005,Moll:APJ2011}.   Some researchers~\cite{Dar:PD2001,Debliquy:PP2005,Carati:JT2006}  employ logarithmic-binned shells to calculate shell-to-shell energy transfers, while others~\cite{Alexakis:PRE2005,Moll:APJ2011} used linearly-binned shells.  Moll {\it et al.}~\cite{Moll:APJ2011} showed that during the dynamo growth for $\mathrm{Pm} =1$, energy transfers take place from large-scale velocity field to small-scale magnetic field.   For $\mathrm{Pm}=1$, the steady state behaviour is that the magnetic-to-magnetic energy transfer is forward and local; there is a strong energy transfer from the large-scale velocity field to the large-scale magnetic field.  We find some of these behaviour repeating in dynamo, specially in small-Pm dynamo, when the magnetic energy has become significantly large.

For the study of large-Pm dynamo ($\mathrm{Pm} =20$) or small-scale dynamo, Kumar {\it et al.}~\cite{Kumar:EPL2013} employed the formalism of Dar {\it et al.}~\cite{Dar:PD2001}, Verma~\cite{Verma:PR2004}, and Debliquy {\it et al.}~\cite{Debliquy:PP2005} to compute energy transfers during the magnetic energy growth. For SSD, Kumar {\it et al.}~\cite{Kumar:EPL2013}  observed that the growth of the small-scale magnetic field is due to the nonlocal energy transfers from small wavenumber velocity modes to large wavenumber magnetic modes. In the present paper, we compute the energy transfers, specially the energy fluxes and shell-to-shell energy transfers, for dynamo with $\mathrm{Pm} <1$. 

The paper is organized as follows: In Sec.~\ref{sec:theory}, we present governing MHD equations, and the formalism for the calculations of the energy fluxes and shell-to-shell energy transfer rates. Details of our numerical simulation are presented in Sec.~\ref{sec:simulation}. In Sec.~\ref{sec:lsd_pm_0.2}, we present results of the forced MHD simulation for $\mathrm{Pm} =0.2$.  The results of the forced MHD simulation for $\mathrm{Pm} =20$ is presented in Sec.~\ref{sec:ssd_pm_20}. In Sec.~\ref{sec:dec_sim}, we present the results of the decaying simulation for $\mathrm{Pm} =20$ and $\mathrm{Pm} =0.2$. Finally, in Sec.~\ref{sec:conclude}, we summarize our simulation results.     
     
\section{Theoretical background} 
\label{sec:theory}
The governing equations for the dynamo process are the incompressible MHD equations~\cite{Verma:PR2004}
\begin{eqnarray}
\partial_{t}\mathbf{u}+ (\mathbf{u} \cdot \nabla) \mathbf{u} & = & -\nabla \left(\frac{p}{\rho}\right)+ \frac{\mathbf{J} \times \mathbf{b}}{\rho}
+ \nu \nabla^{2}\mathbf{u} + \mathbf{F}, \label{eq:MHD_vel}\\
\partial_{t}\mathbf{b}+ (\mathbf{u} \cdot \nabla) \mathbf{b} & = & (\mathbf{b} \cdot \nabla) \mathbf{u}+ \eta \nabla^{2}\mathbf{b}, \label{eq:MHD_mag}\\
\nabla \cdot \mathbf{u} & = & 0, \\ \label{eq:div_u_0}
\nabla \cdot \mathbf{b} & = & 0,  \label{eq:div_b_0}
\end{eqnarray}
where $\mathbf{u}$ and $\mathbf{b}$ are the velocity and magnetic fields, respectively, $p$ is the thermal pressure, $\mathbf{J} = \nabla \times \mathbf b$ is the current density, $\rho$ is the constant fluid density, and $\mathbf{F}$ is the external force field.   To understand the energy transfers between velocity and magnetic fields, we compute energy fluxes and shell-to-shell energy transfer rates using the method proposed by Dar {\it et al.}~\cite{Dar:PD2001}, Verma~\cite{Verma:PR2004}, and Debliquy {\it et al.}~\cite{Debliquy:PP2005}. The energy flux from the region  $X$ (of wavenumber space) of $\alpha$ field  to the region  $Y$ of $\beta$ field is defined as  
\begin{eqnarray}
\Pi^{\alpha,X}_{\beta,Y} = \displaystyle\sum_{\mathbf{k} \in Y} \displaystyle\sum_{\mathbf{p} \in X} S^{\beta \alpha} (\mathbf{k} | \mathbf{p} |\mathbf{q}), \label{eq:flux}
\end{eqnarray}
where $S^{\beta \alpha} (\mathbf{k}| \mathbf{p}| \mathbf{q})$ represents energy transfer rate from mode $\mathbf{p}$ of $\alpha$ field to mode $\mathbf{k}$ of $\beta$ field with the mode $\mathbf{q}$ acting as a mediator.  Here the triadic modes ($ {\mathbf {k,p,q}}$) satisfy a condition  $\mathbf{k} + \mathbf{p} + \mathbf{q} = 0$. For example, energy transfer rate from $\mathbf{u} (\mathbf{p})$ to $\mathbf{b} (\mathbf{k})$ is 
\begin{eqnarray}
S^{bu} (\mathbf{k} | \mathbf{p} | \mathbf{q}) = \Im ([\mathbf{k} \cdot \mathbf{b} (\mathbf{q})][\mathbf{b} (\mathbf{k}) \cdot \mathbf{u} (\mathbf{p})]),\label{eq:mode-to-mode}
\end{eqnarray}
where $\Im$ denotes the imaginary part of the argument, and $\mathbf{b} (\mathbf{q})$ acts as a mediator.

In MHD turbulence there are six energy fluxes: $\Pi^{u<}_{u>}(k_0)$, $\Pi^{u<}_{b>}(k_0)$, $\Pi^{b<}_{b>}(k_0)$, $\Pi^{b<}_{u>}(k_0)$, $\Pi^{u<}_{b<}(k_0)$, and $\Pi^{u>}_{b>}(k_0)$. Here $<$ and $>$ represent the modes residing inside and outside the sphere of radius $k_0$, respectively. Dar {\it et al.}~\cite{Dar:PD2001} and Verma~\cite{Verma:PR2004} proposed formulas to compute these fluxes. For example, the energy flux from inside of the $u$-sphere of radius $k_0$ to outside of the $b$-sphere of the same radius is  
\begin{eqnarray}
\Pi^{u<}_{b>}(k_0) = \displaystyle\sum_{|\mathbf{k}| > k_0} \displaystyle\sum_{|\mathbf{p}| < k_0} S^{b u} (\mathbf{k} | \mathbf{p} | \mathbf{q}). \label{eq:flux_ub}
\end{eqnarray}

In addition, Dar {\it et al.}~\cite{Dar:PD2001} also formulated the shell-to-shell energy transfer rates for MHD turbulence that provide further insights into the energy transfers in wavenumber space. In MHD, we discuss three kinds of shell-to-shell energy transfer rates~\cite{Dar:PD2001,Verma:PR2004} -- from velocity to velocity field ($U2U$), from magnetic to magnetic ($B2B$), and from velocity to magnetic ($U2B$). The shell-to-shell energy transfer from the $m$-th shell of $\alpha$ field to the $n$-th shell of $\beta$ field is defined as~\cite{Debliquy:PP2005, Dar:PD2001, Verma:PR2004} 
\begin{eqnarray}
T^{\beta,\alpha}_{n,m} = \displaystyle\sum_{\mathbf{k} \in n} \displaystyle\sum_{\mathbf{p} \in m} S^{\beta \alpha} (\mathbf{k} | \mathbf{p} | \mathbf{q}). \label{eq:shell-to-shell}
\end{eqnarray}
As an illustration, the shell-to-shell energy transfer rate from the $m$-th shell of $u$ field to the $n$-th shell of $b$ field is  given by 
\begin{eqnarray}
T^{b,u}_{n,m} = \displaystyle\sum_{\mathbf{k} \in n} \displaystyle\sum_{\mathbf{p} \in m} S^{b u} (\mathbf{k} | \mathbf{p} | \mathbf{q}). \label{eq:shell-to-shell_ub}
\end{eqnarray}

In the next section, we describe the numerical method adopted for dynamo simulations.

\section{Details of numerical simulation} 
\label{sec:simulation}

We perform our numerical simulations using a pseudo-spectral code {\em Tarang}~\cite{Verma:Pramana2013}. The simulations have been performed in a three-dimensional box of size $(2\pi)^3$ with periodic boundary conditions in all the three directions. The grid size for our simulations is $1024^3$. We employ Runge-Kutta  fourth order (RK4) scheme for time integration,  $2/3$ rule for dealiasing, and CFL criterion for choosing $\Delta t$.  We performed extensive validation tests on Tarang including the grid-independence test~\cite{Verma:Pramana2013}.  Refer to Reddy and Verma~\cite{Reddy:PF2014} for the grid-independence test.

We perform several sets of forced and decaying dynamo simulations for $\mathrm{Pm} =0.2$ and $20$.   The parameters for these runs are listed in Table~\ref{table:sim_para}.   For the forced simulations, we apply nonhelical random forcing to the velocity field in a wavenumber band $k =[2, 4]$ such that the supply rate of the kinetic energy  is a constant.   We also perform a low resolution dynamo simulation for $\mathrm{Pm} =0.2$, which will be discussed in Sec.~\ref{sec:low_resolution}. We could not perform dynamo simulations for Pm lower than 0.2 due to large grid and  computation time requirements.

By applying the same approach as that of Ponty {\it et al.}~\cite{Ponty:PRL2005} and Kumar {\it et al.}~\cite{Kumar:EPL2013}, we first perform pure fluid simulations with $\nu =0.002$ for $\mathrm{Pm} =0.2$ and with $\nu =0.01$ for $\mathrm{Pm} =20$ dynamo simulations. The simulations were carried out until the fluid flow becomes statistically steady. After this, the final fluid state and an initial seed magnetic field, whose total magnetic energy of $10^{-4}$ unit is uniformly distributed in a wavenumber band $k= [2, 4]$, are employed as the initial condition for the dynamo run.  Since every mode has an equal energy, $E_b(k) \sim k^2$ for $k= [2, 4]$, thus the seed magnetic field is active at the large scales.  We continue our dynamo simulation for approximately $20$ eddy turnover time. 

We also perform decaying dynamo simulations for $\mathrm{Pm} =0.2$ and $20$ by turning off the forcing. We observe that the evolution of the decaying dynamo differs significantly from the forced one; these differences will be discussed in Sec.~\ref{sec:dec_sim}.

In the present paper, we compute the energy fluxes and shell-to-shell energy transfers for dynamo.  For these computations,  the wavenumber space is divided into $19$ shells.   The first three shell radii are 2, 4, and 8, respectively, while the  last  two shell radii are $341$ ($=512\times2/3$) and $170.5$ ($=(512\times 2/3)/2$), respectively.  The factor 2/3 arises due to dealiasing.  The remaining shells are  logarithmically binned, which yields the shell radii as:  $2.0$, $4.0$, $8.0$, $9.8$, $12.0$, $14.8$, $18.1$, $22.2$, $27.2$, $33.4$, $40.9$, $50.2$, $61.5$, $75.4$, $92.5$, $113.4$, $139.0$, $170.5$, and $341.0$.


\begin{table}[htbp]
\caption{Parameters of the simulations: kinematic viscosity ($\nu$), magnetic diffusivity ($\eta$), magnetic Prandtl number ($\mathrm{Pm}$),  the final $u_\mathrm{rms}$, the final $b_\mathrm{rms}$, velocity integral length scale $L_u$ ($=2\pi \int k^{-1} E_u(k) dk / \int E_u(k) dk$),  magnetic integral length scale $L_b$ ($=2\pi \int k^{-1} E_b(k) dk / \int E_b(k) dk$), kinetic Reynolds number $\mathrm{Re}$ ($=u_\mathrm{rms}L_u / \nu$), and magnetic Reynolds number $\mathrm{Rm}$ ($=u_\mathrm{rms}L_u / \eta$).    } 
\vspace{0.5pc}
\centering
\begin{tabular}{c c c c c c c c c c c}
\hline 
\hline  
Run & grid & $\nu$ & $\eta$ & $\mathrm{Pm}$ & $u_\mathrm{rms}$ & $b_\mathrm{rms}$ & $L_u$  & $L_b$  & $\mathrm{Re}$ & $\mathrm{Rm}$ \\ [1ex]
\hline 
FluidA & $1024^3$ & $0.01$ & $-$ & $-$ & $5.09$ & $-$ & $1.31$ & $-$ & $666$ & $-$ \rule{0pt}{3ex}   \\

FluidB & $1024^3$  & $0.002$ & $-$ & $-$ & $1.83$ & $-$ & $1.45$ & $-$ & $1327$ & $-$\\

FluidC &  $256^3$  & $0.02$ & $-$ & $-$ & $2.10$ & $-$ & $1.77$ & $-$ & $185$ & $-$\\ 
 
DynamoA & $1024^3$  & $0.01$ & $0.0005$ & $20$ & $0.86$ & $0.80$ & $1.67$ & $0.56$ & $143$ & $2860$\\

DynamoB & $1024^3$  & $0.002$ & $0.01$ & $0.2$ & $1.38$ & $0.02$ & $1.31$ & $0.76$ & $905$ & $181$\\ 

DynamoC & $256^3$  & $0.02$ & $0.1$ & $0.2$ & $1.73$ & $0.86$ & $1.79$ & $3.11$ & $155$ & $31$\\ 

Dynamo DecayA & $1024^3$  & $0.01$ & $0.0005$ & $20$ & $\rightarrow 0$ & $\rightarrow 0$ & $2.85$ & $1.67$ & $20$ & $400$\\
 
Dynamo DecayB & $1024^3$  & $0.002$ & $0.01$ & $0.2$ & $\rightarrow 0$ & $\rightarrow 0$ & $1.27$ & $0.93$ & $438$ & $88$\\[1ex]
\hline
\hline
\end{tabular}
\label{table:sim_para} 
\end{table}

Note that we use nondimensionalized equations, hence the time of our equation is nondimensional.  The integral length  is of the order of unity, so one eddy turnover time is $1/u_\mathrm{rms}$ is of the order of unity.

In the next section, we describe the results of our dynamo simulations.

\section{Dynamo simulation for $\mathrm{Pm} =0.2$}
\label{sec:lsd_pm_0.2}

In this section, we present results of the dynamo simulation for $\mathrm{Pm} =0.2$. In the first subsection, we describe the growth of the kinetic and magnetic energies with time. In the subsequent subsections, we discuss the energy fluxes and shell-to-shell energy transfers during the growth phase of the magnetic energy. In the last subsection, we also present results of the dynamo simulation for $\mathrm{Pm} =0.2$ with a lower resolution and a lower $\mathrm{Rm}$.
 
\subsection{Growth of kinetic and magnetic energy}
\label{sec:lsd_en_growth}

\begin{figure}[htbp]
\centering
\includegraphics[scale=0.65]{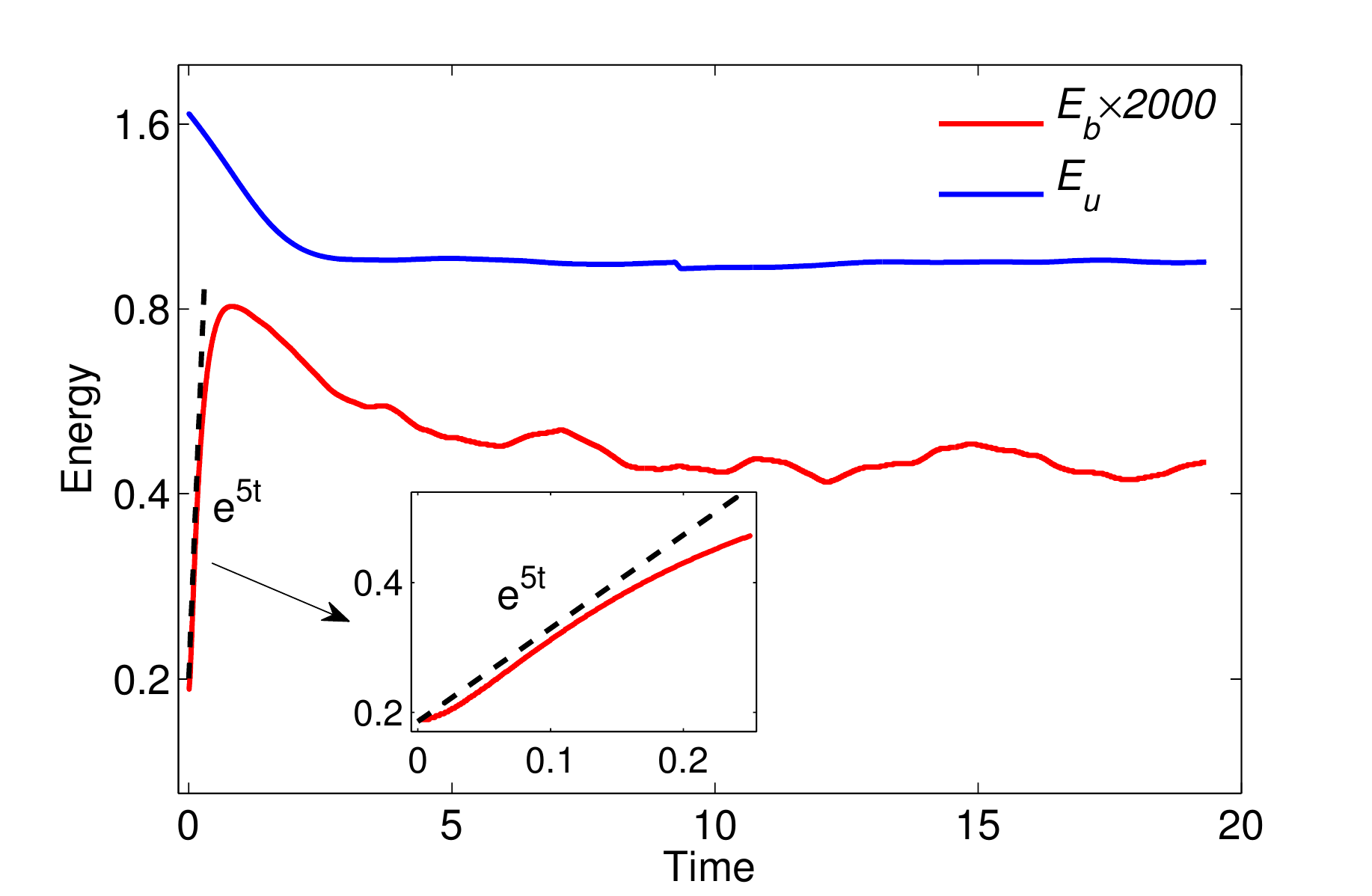}  
\caption{For dynamo with  $\mathrm{Pm} =0.2$: Evolution of kinetic energy ($E_u$) and magnetic energy ($E_b$) with time, where the vertical axis is logarithmic. The subfigure exhibits a zoomed view of the growth phase of the magnetic energy. In the initial stages of the simulation, the magnetic energy grows exponentially with time ($E_b(t) \sim e^{5t}$). The kinetic and magnetic energies saturate quickly. In the final stages of the simulation, $E_u/E_b \approx 4000$, so $E_b$ is multiplied by $2000$ to fit in the same plot.}
\label{fig:lsd_kin_mag}
\end{figure}

In Fig.~\ref{fig:lsd_kin_mag}, we exhibit the growth of the kinetic energy and magnetic energy with time. A zoomed view of the growth phase of the magnetic energy is shown in a subfigure of Fig.~\ref{fig:lsd_kin_mag}. The magnetic energy grows exponentially from $t=0$ to $0.25$ as $E_b(t) = E_b(t=0) e^{\gamma t}$ with $\gamma \approx 5$. After reaching the peak, $E_b$ decreases to approximately half of its peak value, and then, at $t \approx 10$, it saturates.   The saturated values of $E_u$ and $E_b$ continue till $t \approx 20$ eddy turnover time, which is the maximum time of our simulation.   We expect $E_b$ to continue at the saturated value at later time as well.   The rapid growth of $E_b$ appears to negate a possibility of transient growth, but a combination of rapid growth and decay before saturation also strengthens this possibility. These issues (non-normal growth) needs to be investigated in detail, but we avoid its discussion in the present paper since our focus is on the energy transfers in dynamo.  

Figure~\ref{fig:lsd_kin_mag} also illustrates that the kinetic energy $E_u$ decreases during the growth phase of $E_b$, and then it saturates along with $E_b$.  The kinetic Reynolds number $\mathrm{Re}= u_{\mathrm{rms}}L_u/\nu$ and the magnetic Reynolds number $\mathrm{Rm} = u_{\mathrm{rms}}L_u/\eta$  for the steady state are $905$ and $181$, respectively.  Here the velocity integral length scale $L_u =2\pi \int k^{-1} E_u(k) dk / \int E_u(k) dk$ and the magnetic integral length scale $L_b =2\pi \int k^{-1} E_b(k) dk / \int E_b(k) dk$.

It is interesting to observe that the saturation value of magnetic energy is $4000$ times smaller than that of the kinetic energy, or $E_b/E_u \approx 1/4000$, which is a strong deviation from the typical equipartition $E_b/E_u \approx 1$ observed in many numerical simulations and in solar wind.  The very small value of $E_b/E_u$  in our $1024^3$ run was indeed a surprising result that led us to investigate the simulations in much greater detail.   Many researchers had reported $E_b/E_u$ very different from unity near the dynamo transition in numerical simulations  (see e.g., Glatzmaier and Roberts~\cite{Glatzmaier:PEPI1995}, Yadav {\it et al.}~\cite{Yadav:EPL2010}, Cattaneo and Vainshtein~\cite{Cattaneo:APJ1991}) as well as in the von Karman Sodium (VKS) experiment~\cite{Monchaux:PRL2007}. In a numerical simulation of convective dynamo, Glatzmaier and Roberts~\cite{Glatzmaier:PEPI1995} observed $E_b/E_u \approx 10^3$. For a nonhelical dynamo simulation  with $\mathrm{Pm =1}$, Meneguzzi {\it et al.}~\cite{Meneguzzi:PRL1981} observed that $E_b/E_u \approx 0.1$, and Yadav {\it et al.}~\cite{Yadav:EPL2010} found the ratio to vary from $10^{-2}$ to $3$.  In another dynamo simulation for $\mathrm{Pm} =0.5$, Mishra {\it et al.}~\cite{Mishra:EPL2013} observed the ratio to be $1/16$. In a recent simulation of convective dynamo for $\mathrm{Pm} =0.2$, Guervilly {\it et al.}~\cite{Guervilly:PRE2015} observed that $E_b/E_u \approx 10^{-2}$.   In VKS experiment~\cite{Monchaux:PRL2007,Aumaitre:CRP2008}, $E_b/E_u$ was observed to be much less than unity.  

The divergence from equipartition is possibly due to the multiple attractors present in the chaotic regime~\cite{Yadav:EPL2010,Yadav:PRE2012,Aumaitre:CRP2008}.  It is conjectured that the equipartition is probably a robust property of the attractor of the fully-developed turbulence. Note however that the attractor of the fully-developed turbulence is not ergodic~\cite{Lesieur:book1997}.  The dynamo with $\mathrm{Pm}=0.2$ simulated by us is near the onset.  In the steady state, the magnetic Reynolds number $\mathrm{Rm} = 181$.  When we decreased $\mathrm{Rm}$ to $120$ by increasing the magnetic diffusivity, the magnetic energy decays to zero. Hence, we show that our dynamo is near the onset.   The fully-developed turbulence regime for $\mathrm{Pm}=0.2$ requires much  higher resolution than $1024^3$ as well as much longer computation time.

For low magnetic Prandtl number, the dynamo regime is turbulent, i.e., $\mathrm{Re} \gg 1$. For such flows, Cattaneo and Vainshtein~\cite{Cattaneo:APJ1991} and Verma~\cite{Verma:CS2002,Verma:PR2004} argued that the magnetic energy growth time scale is 
\begin{eqnarray}
\tau_\mathrm{turb} = \frac{L_{u}^2}{\eta_\mathrm{turb}} = \frac{L_u}{u_\mathrm{rms}}, \label{eq:tau_turb_uqn}
\end{eqnarray}
which is an eddy turnover time because $\eta_\mathrm{turb} = u_\mathrm{rms}L_u$. This is the reason why in our simulation the magnetic energy saturates in several eddy turnover time.

  
\begin{figure}[htbp]
\centering
\includegraphics[scale=0.55]{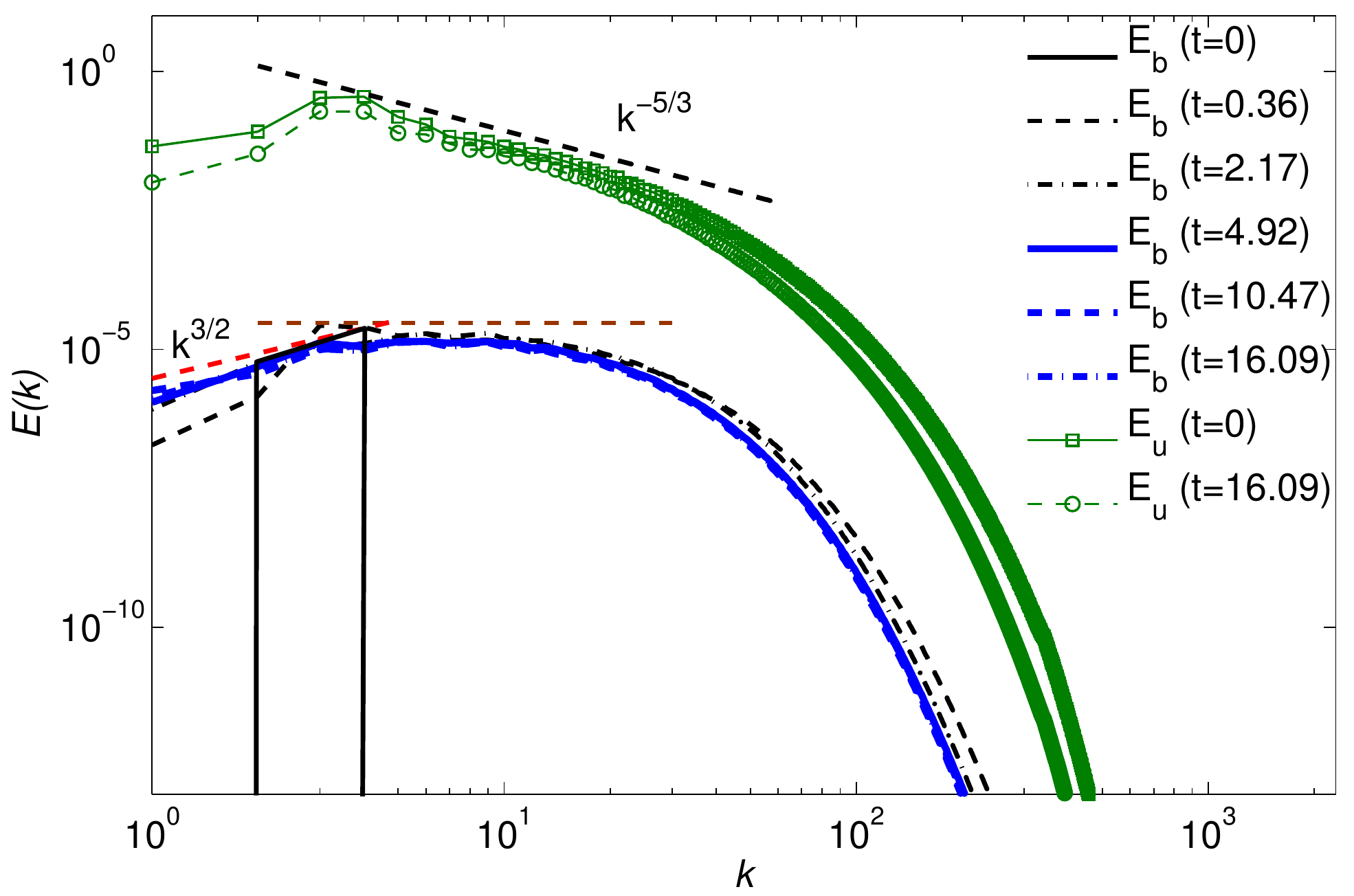}  
\caption{For dynamo with  $\mathrm{Pm} =0.2$: Evolution of kinetic ($E_u(k)$) and magnetic ($E_b(k)$) energy spectra with time. The magnetic energy is flat for intermediate $k$. The kinetic energy follows the Kolmogorov scaling ($E_u(k) \sim k^{-5/3}$).}
\label{fig:Ebk_LS}
\end{figure}

The time evolution of the kinetic energy spectrum ($E_u(k)$) and the magnetic energy spectrum ($E_b(k)$) are shown in Fig.~\ref{fig:Ebk_LS}.    The magnetic field applied in the wavenumber band $k =[2, 4]$ at $t=0$ spreads out rapidly to the whole wavenumber space. The magnetic energy spectrum saturates in several eddy turnover.  Under the steady state, the magnetic energy spectrum is flat for the intermediate wavenumbers.  The small wavenumber region appears to exhibit a narrow band with $E_b(k) \sim k^{3/2}$, but this spectrum is inconclusive.  The kinetic energy spectrum does not vary significantly over time, and it maintains the Kolmogorov scaling ($E_u(k) \sim k^{-5/3}$).  


\begin{figure}[htbp]
\centering
\includegraphics[scale=0.6]{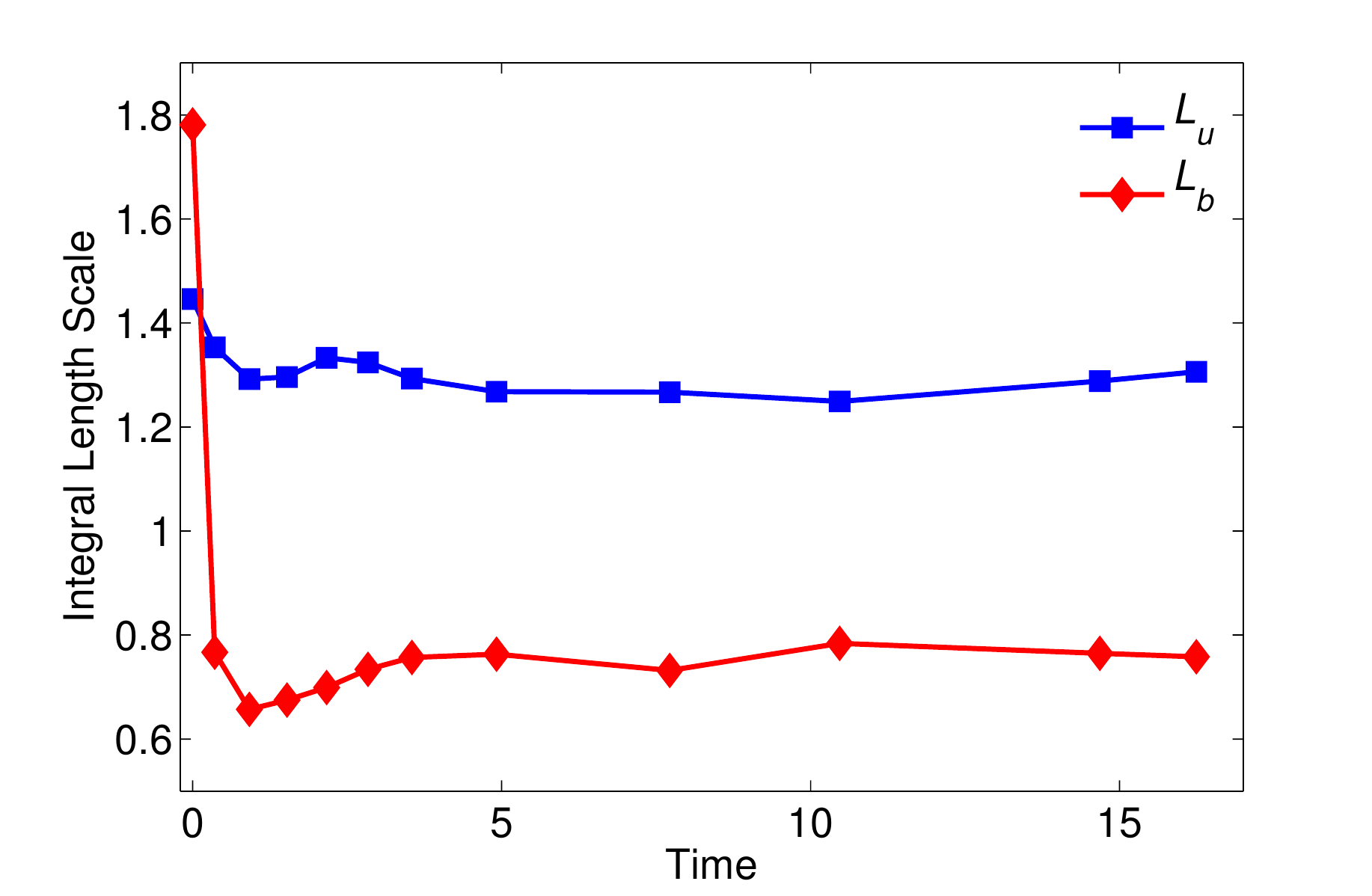} 
\caption{For dynamo with  $\mathrm{Pm} =0.2$: Plots exhibiting  time dependence of the velocity and magnetic integral length scales ($L_u, L_b$). $L_u$ saturates quickly, while $L_b$ first decreases sharply and then saturates.}
\label{fig:lsd_integ_len_time}
\end{figure}

In Fig.~\ref{fig:lsd_integ_len_time}, we plot the velocity and magnetic integral length scales as a function of time.  In the early stages,  $L_b > L_u$, but at a later time, $L_b$ decreases and then it saturates to approximately 0.8 with  asymptotic $L_u/L_b \approx 1.5$.    The velocity integral length scale $L_u$ does not change significantly with time and saturates quickly to a constant value of around 1.3.  


In the next subsection, we will discus the energy fluxes in dynamo with $\mathrm{Pm} =0.2$.

\subsection{Energy Fluxes}

\begin{figure}[htbp]
\centering
\includegraphics[scale=0.65]{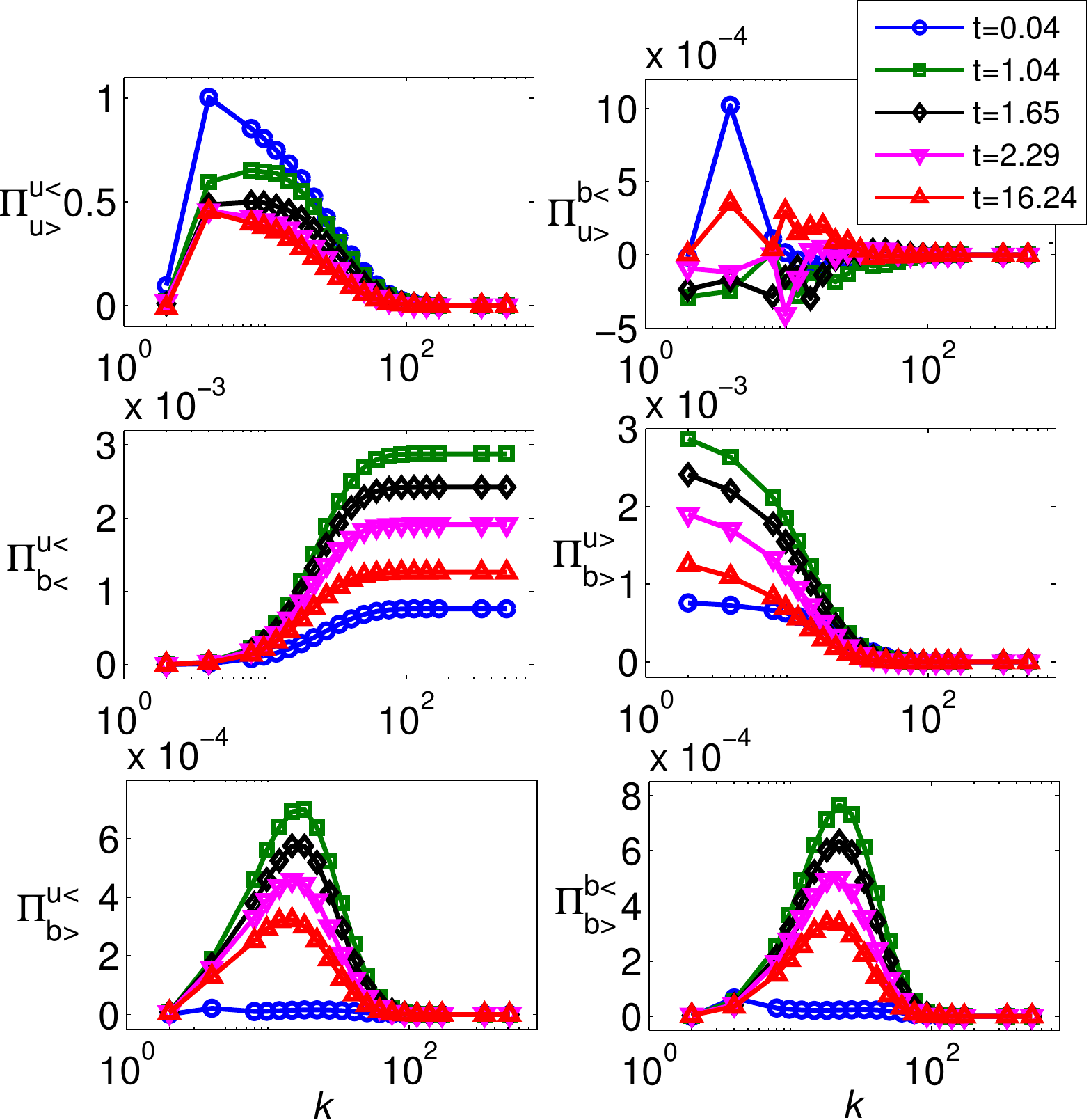}  
\caption{For dynamo with  $\mathrm{Pm} =0.2$: Plots of energy fluxes $\Pi^{u<}_{u>}$, $\Pi^{b<}_{u>}$, $\Pi^{u<}_{b<}$, $\Pi^{u>}_{b>}$, $\Pi^{u<}_{b>}$, and $\Pi^{b<}_{b>}$ vs. $k$. All the energy fluxes, except $\Pi^{b<}_{u>}$, are positive. The energy flux $\Pi^{u<}_{u>}$ (from velocity to velocity) is the most dominant among all the fluxes, but $\Pi^{u<}_{b<}$ is the most dominant flux for the growth of $E_b$.}
\label{fig:lsd_flux}
\end{figure}

To investigate the energy transfers during the  growth of the magnetic energy, we now focus on the energy flux computations~\cite{Dar:PD2001,Verma:PR2004}. A quantitative analysis of various energy fluxes of  MHD is one of the main features of this work. In Fig.~\ref{fig:lsd_flux}, we present energy fluxes computed during the $E_b$ growth for $\mathrm{Pm} =0.2$. The energy fluxes from the inner $u$-sphere to the outer $u$-sphere ($\Pi^{u<}_{u>}$), the inner $u$-sphere to the inner $b$-sphere ($\Pi^{u<}_{b<}$), the inner $u$-sphere to the outer $b$-sphere ($\Pi^{u<}_{b>}$),  the inner $b$-sphere to the outer $b$-sphere ($\Pi^{b<}_{b>}$), and the outer $u$-sphere to the outer $b$-sphere ($\Pi^{u>}_{b>}$) are all positive, whereas the energy flux from the inner $b$-sphere to the outer $u$-sphere ($\Pi^{b<}_{u>}$) takes both positive and negative values. The energy flux $\Pi^{u<}_{u>}$ dominates all the other fluxes, while the other fluxes are of the same order. It is important to note that the energy flux $\Pi^{u<}_{b<}$ is the most dominant flux among those responsible for the growth of the magnetic energy.

The aforementioned energy fluxes are very different from the results reported for $\mathrm{Pm} =20$ by Kumar {\it et al.}~\cite{Kumar:EPL2013}. In their numerical simulation, they observed that the energy flux $\Pi^{u<}_{b>}$ dominates all the other fluxes, i.e., the magnetic energy growth is mainly due to the energy transfers from the velocity field at small wavenumbers to the magnetic field at large wavenumbers. It is evident from Fig.~\ref{fig:lsd_flux} that the initial energy fluxes $\Pi^{u<}_{b<}$, $\Pi^{u>}_{b>}$, $\Pi^{u<}_{b>}$, and $\Pi^{b<}_{b>}$ increase rapidly (at $t=1.04$),  then decrease, and finally saturate (see the final fluxes at $t=16.24$). The energy fluxes $\Pi^{u<}_{b>}$ and $\Pi^{b<}_{b>}$ peak at the wavenumber $k \approx 20$ at all time, which indicates that the scale at which magnetic energy grows does not change with time.

In the next subsection, we will discus the shell-to-shell energy transfers in dynamo with $\mathrm{Pm} =0.2$.


\subsection{Shell-to-shell energy transfers}

\begin{figure}[htbp]
\centering
\includegraphics[scale=1.0]{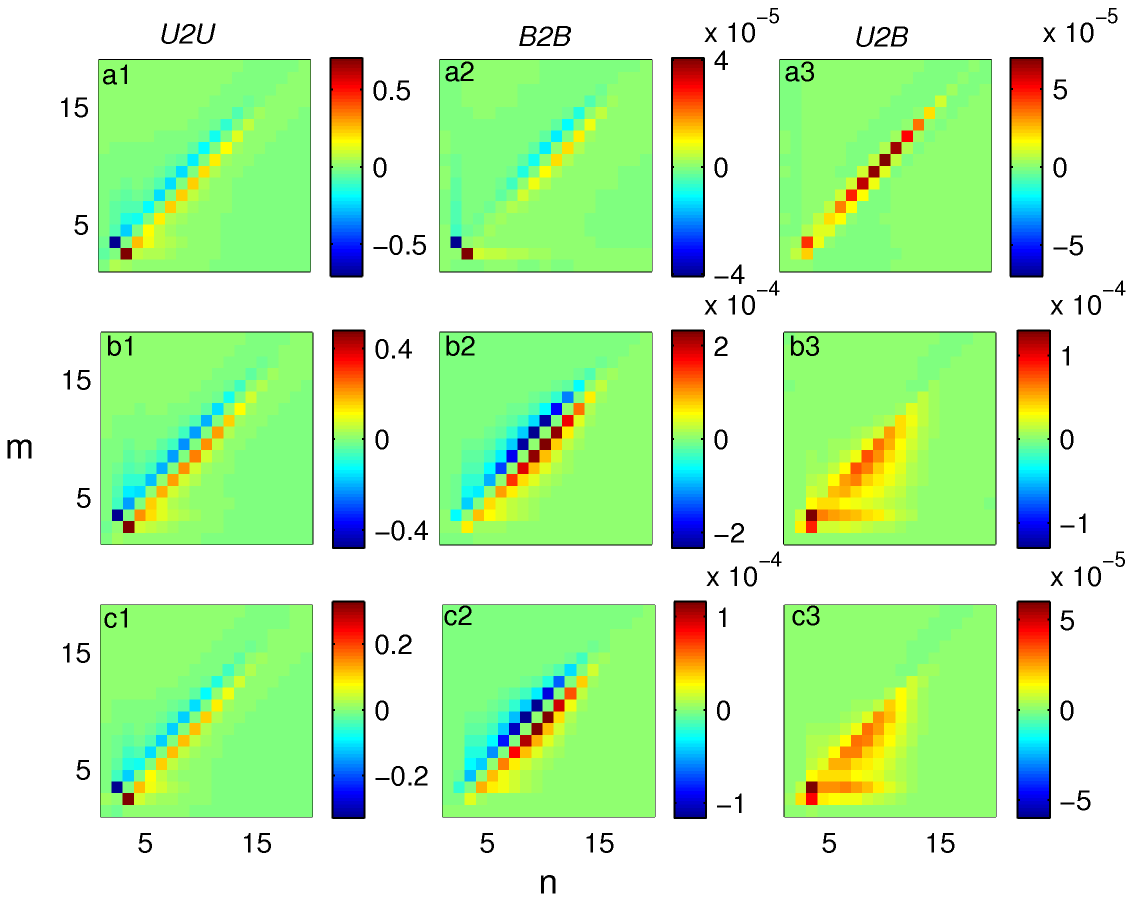}  
\caption{For dynamo with  $\mathrm{Pm} =0.2$: Shell-to-shell energy transfer rates, $U2U$ (velocity to velocity), $B2B$ (magnetic to magnetic), and $U2B$ (velocity to magnetic) at  $t=0.04$ ((a1)-(a3)),  $t=0.47$ ((b1)-(b3)),  and $t=16.24$ ((c1)-(c3)). Here the horizontal axes represent the receiver shells, whereas the vertical axes represent the giver shells. The $U2U$, $B2B$, and $U2B$ energy transfers are local and forward.}
\label{fig:lsd_shell_uu_bb_ub}
\end{figure}

The energy fluxes provide a broader view of the energy transfer mechanisms. To obtain an elaborate picture of the energy transfers in dynamo for $\mathrm{Pm} =0.2$, we compute the shell-to-shell energy transfer rates~\cite{Dar:PD2001, Verma:PR2004}. In Fig.~\ref{fig:lsd_shell_uu_bb_ub}, we present velocity to velocity ($U2U$), magnetic to magnetic ($B2B$), and velocity to magnetic ($U2B$) energy transfers for $\mathrm{Pm} =0.2$. The $U2U$ and $B2B$ energy transfers are forward and local, i.e., the energy transfers occur from smaller wavenumbers to larger wavenumbers, and it is predominantly among neighboring wavenumber shells. In the early stages of the simulation (shown in Fig.~\ref{fig:lsd_shell_uu_bb_ub}(a2) at $t=0.04$), the $B2B$ energy transfer is concentrated at small wavenumbers, but later it spreads to larger wavenumbers. The  initial $B2B$ energy transfers at smaller wavenumbers is due to the initial  seed magnetic field applied at smaller wavenumbers ($k =[2, 4]$). Later, the growth of the magnetic energy takes place at intermediate wavenumbers, hence the $B2B$ transfer develops at intermediate wavenumbers. 

\begin{figure}[htbp]
\centering
\includegraphics[scale=0.5]{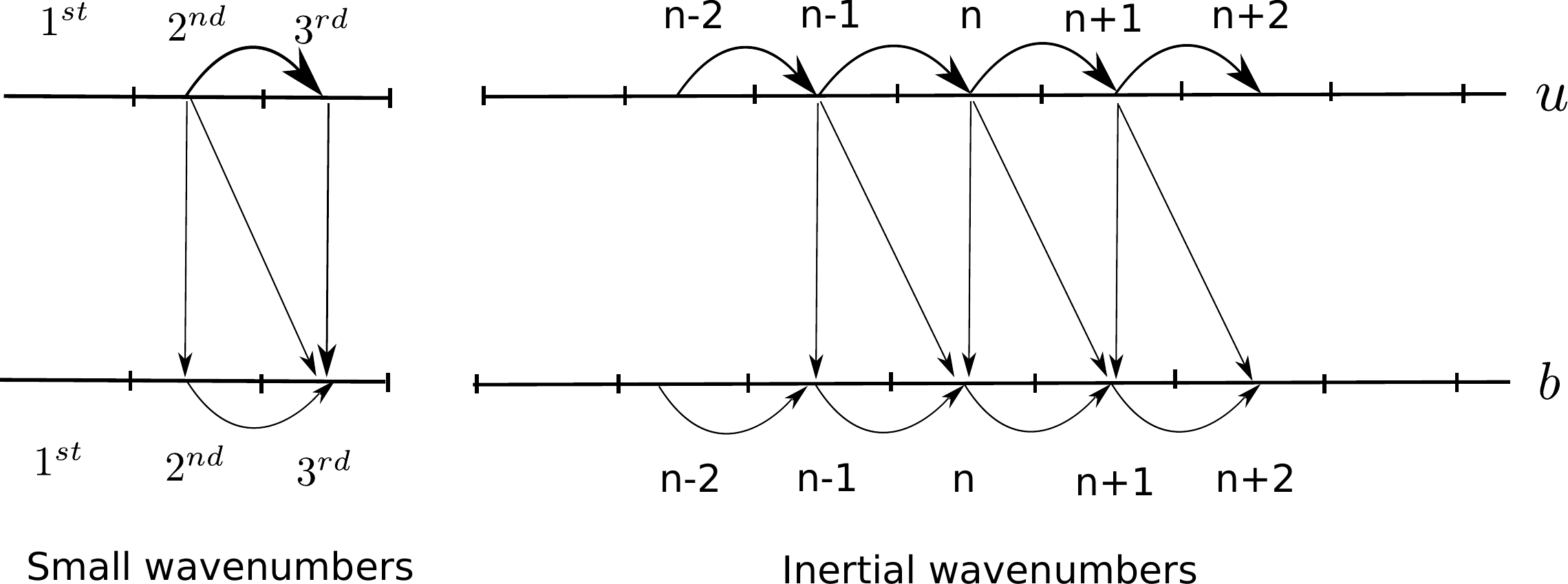} 
\caption{For dynamo with  $\mathrm{Pm} =0.2$: A schematic diagram of shell-to-shell energy transfers among the $u$-shells and the $b$-shells.}
\label{fig:lsd_shell_diag}
\end{figure}

The  $U2B$ energy transfer is positive for all the shells, hence, as expected, the energy transfer takes place from the velocity field to the magnetic field.  
In the initial stages of dynamo, the $U2B$ energy transfer is local, and it is spread at all wavenumbers, indicating a rapid growth of the magnetic energy at all scales (shown in Fig.~\ref{fig:lsd_shell_uu_bb_ub}(a3) at $t=0.04$). Later, we observe  $U2B$ energy transfers from the $3^{rd}$ $u$-shell, which is a forcing shell, to $4^{th}, 5^{th}, 6^{th}, 7^{th}$, and $8^{th}$ $b$-shells.  Hence the $U2B$ energy transfer develops  a weak nonlocal component. The peak of the $U2B$ transfers is concentrated at $3^{rd}$ $u$-shell to the $3^{rd}$ $b$-shell (small wavenumbers) because both the $E_u(k)$ and $E_b(k)$ have significant strength at small wavenumbers. This is consistent with the dominant $\Pi^{u<}_{b<}$ energy flux discussed in the earlier subsection.  In Fig.~\ref{fig:lsd_shell_diag}, we exhibit a schematic representation of the shell-to-shell energy transfers  for dynamo with $\mathrm{Pm} =0.2$. It is evident that $U2U$, $B2B$, and $U2B$ energy transfers are local and forward. 
                 
The dynamo discussed above has certain non-generic properties.  For example, the steady state $E_b/E_u \approx 1/4000$, far away from equipartition of $E_b$ and $E_u$.  This may be because the aforementioned system is near the dynamo transition (refer to the earlier discussion).  To explore whether the $\mathrm{Pm}=0.2$ also admits solution that are closer to equipartition,  we perform another dynamo simulation for $\mathrm{Pm} =0.2$ with a lower resolution and a lower $\mathrm{Rm}$, results of which are discussed in the next subsection.

\subsection{$\mathrm{Pm}=0.2$ dynamo with $\nu =0.02$ and $\eta =0.1$ on $256^3$ grid}
\label{sec:low_resolution}

\begin{figure}[htbp]
\centering
\includegraphics[scale=0.65]{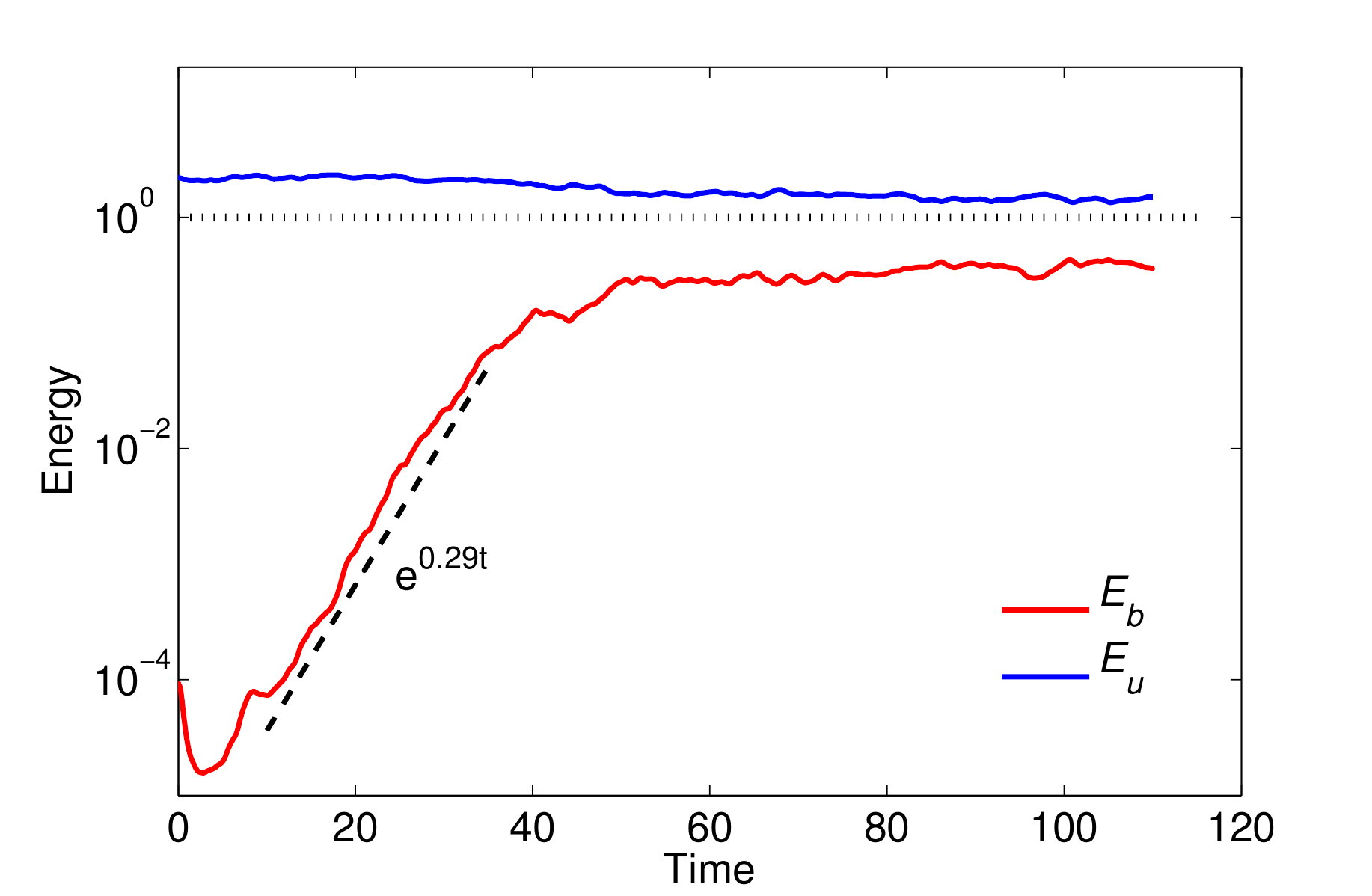} 
\caption{For dynamo with $\mathrm{Pm} =0.2$ ($\nu =0.02, \eta =0.1$): Evolution of kinetic energy ($E_u$) and magnetic energy ($E_b$) with time. In the initial stages of the simulation, the magnetic energy grows exponentially with time ($E_b(t) \sim e^{0.29t}$).}
\label{fig:lsd_n_256_kin_mag}
\end{figure}

We perform a dynamo simulation for $\mathrm{Pm} =0.2$ on $256^3$ grid with $\nu =0.02$ and $\eta =0.1$ (10 times larger than that for $1024^3$ run).  Note that $\nu$ and $\eta$ for this run differs from those for the $1024^3$ run.  Similar to the earlier approach, for the dynamo simulation, we take a statistically steady fluid flow (with $\nu=0.02$) and apply a small seed magnetic field in the wavenumber band $k=[2,4]$. The simulation parameters are listed in Table~\ref{table:sim_para}. The evolution of kinetic energy and magnetic energy with time is shown in Fig.~\ref{fig:lsd_n_256_kin_mag}. The magnetic energy grows as $e^{0.29t}$, and saturates near about $t=50$ eddy turnover time. The magnetic Reynolds number under the steady state is approximately 31, hence we expect the dynamo to be near the onset.  In the final stages of the simulation, we observe the ratio $E_b/E_u \approx 1/5$, which is quite different from $E_b/E_u \approx 1/4000$ observed in the dynamo simulation for $1024^3$ run.  As described in Sec.~\ref{sec:lsd_en_growth}, this is due to the presence of different chaotic attractors near the dynamo transition. We also remark that the growth of the magnetic energy differs for the two different resolutions.

\begin{figure}[htbp]
\centering
\includegraphics[scale=0.03]{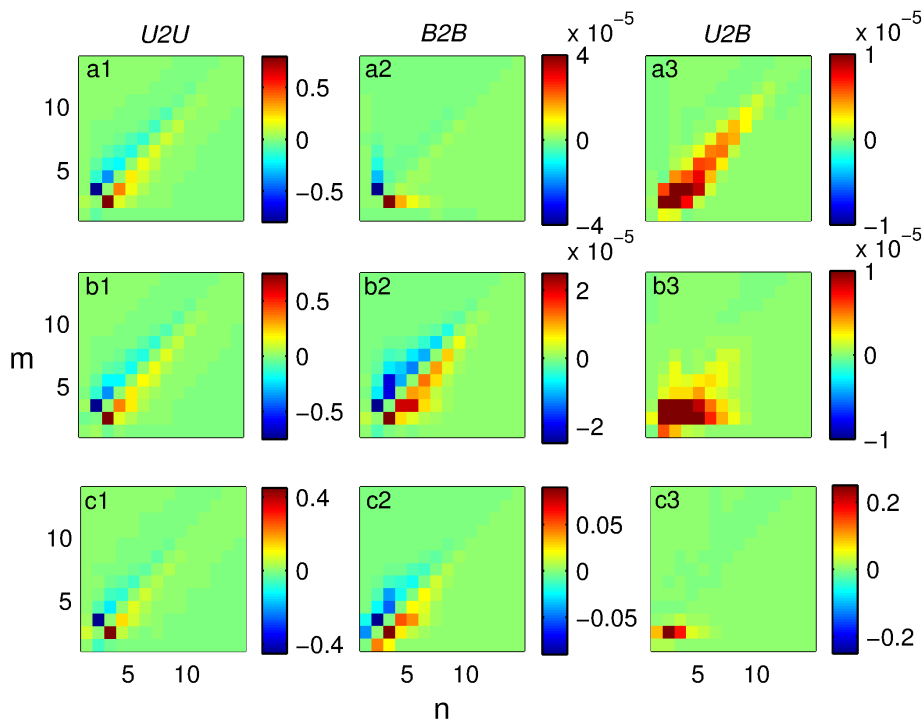} 
\caption{For dynamo with $\mathrm{Pm} =0.2$ ($\nu =0.02, \eta =0.1$): Shell-to-shell energy transfer rates, $U2U$ (velocity to velocity), $B2B$ (magnetic to magnetic), and $U2B$ (velocity to magnetic) at  $t=0.05$ ((a1)-(a3)),  $t=0.5$ ((b1)-(b3)),  and $t=110$ ((c1)-(c3)). The notation is the same as in Fig.~\ref{fig:lsd_shell_uu_bb_ub}. The $U2U$, $B2B$, and $U2B$ energy transfers are local and forward.}
\label{fig:lsd_n_256_shell_uu_bb_ub}
\end{figure}

In Fig.~\ref{fig:lsd_n_256_shell_uu_bb_ub}, we present shell-to-shell energy transfers during the dynamo growth. For $256^3$ simulation, the wavenumber space is divided into $14$ logarithmically binned shells. The shell radii are:  $2.0$, $4.0$, $6.0$, $7.1$, $8.4$, $9.9$, $11.7$, $13.8$, $16.3$, $19.3$, $22.8$, $26.9$, $31.9$, and $85.0$. We observe that the nature of the energy fluxes are qualitatively similar to what we report in dynamo simulation for $\mathrm{Pm} =0.2$ on $1024^3$ grid.  We also observe that the energy flux $\Pi^{u<}_{b<}$ is the most dominant for the growth of the magnetic energy. In other words, the growth of magnetic energy takes place due to the energy transfers from large-scale velocity field to large-scale magnetic field.  The $U2U$ and $B2B$ energy transfers are forward and local.  The $U2B$ transfers are forward and local, but  with a small nonlocal component.  The nonlocality in $U2B$ transfers for $256^3$ dynamo  appears to be more significant than that for on  a $1024^3$ grid with lower $\nu$ and $\eta$ (see Fig.~\ref{fig:lsd_n_256_shell_uu_bb_ub}(b3)).  This feature may be due to strong correlations between the velocity, magnetic, and forcing fields, and this issue is being investigated in detail.  


The energy fluxes and shell-to-shell energy transfers indicate that the nature of the energy transfers, which is the main result of this paper, remains approximately the same during the growth of dynamo for $\mathrm{Pm} =0.2$ for both $1024^3$ and $256^3$ resolutions.  The initial growth phases differ for these resolutions; this feature needs to be investigated in detail. 

In the next section, we discuss the shell-to-shell energy transfers during the magnetic energy growth in forced MHD simulation for $\mathrm{Pm} =20$ and compare it with the energy transfers for a dynamo with $\mathrm{Pm} =0.2$ (discussed above).

\section{Dynamo simulation for $\mathrm{Pm} =20$}
\label{sec:ssd_pm_20}

We perform dynamo simulations for $\mathrm{Pm}=20$ using the parameters listed in Table~\ref{table:sim_para}.  The initial conditions and forcing schemes are the same as in the case of $\mathrm{Pm} =0.2$.  More details of the simulation and results are discussed in Kumar {\it et al.}~\cite{Kumar:EPL2013}.   Here we contrast the energy transfers in dynamo for $\mathrm{Pm}=0.2$ with those in dynamo for $\mathrm{Pm}=20$ by comparing their shell-to-shell energy transfers.  The contrast between the two dynamos are most evident in the shell-to-shell energy transfers,  which is described here.

\begin{figure}[htbp]
\centering
\includegraphics[scale=0.03]{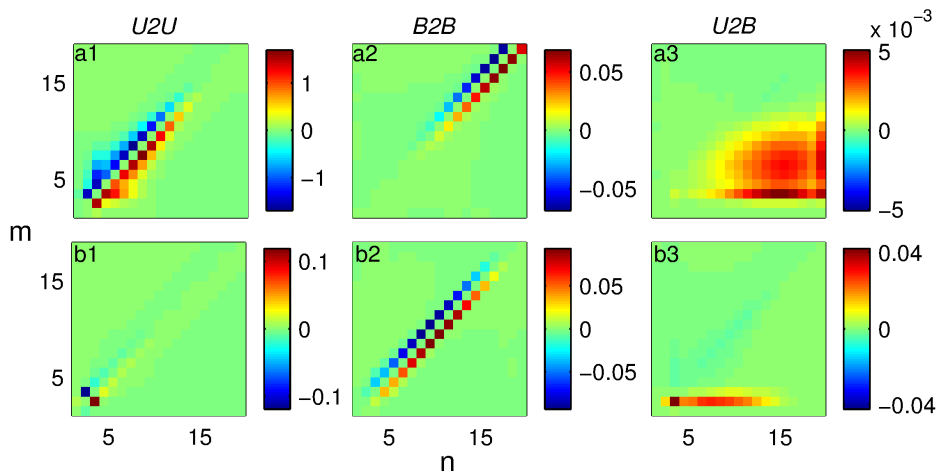}  
\caption{For dynamo with  $\mathrm{Pm} =20$:  Shell-to-shell energy transfer rates at  $t=0.48$ ((a1)-(a3)) and $t=18.73$ ((b1)-(b3)). The notation is the same as in Fig.~\ref{fig:lsd_shell_uu_bb_ub}. The $U2U$ and $B2B$ energy transfers are local and forward, whereas the $U2B$ transfer is nonlocal, forward, and always positive.  Adopted from Kumar {\it et al.}~\cite{Kumar:EPL2013}.}
\label{fig:ssd_shell_uu_bb_ub}
\end{figure}

The $U2U$, $B2B$, and $U2B$ energy transfers for $\mathrm{Pm} =20$ are shown in Fig.~\ref{fig:ssd_shell_uu_bb_ub}. The $U2U$ and $B2B$ energy transfers are forward and local, which are similar to that for $\mathrm{Pm} =0.2$. The $U2B$ transfer is forward and predominantly nonlocal, i.e., from small wavenumber $u$-shells to all the $b$-shells; this nonlocal energy transfer  is responsible for the growth of magnetic energy at small scales.  This is in sharp contrast to the dynamo with $\mathrm{Pm} =0.2$ where $U2B$ transfer is primarily local, and it is dominant for the small wavenumber $u$-shells and $b$-shells.  For dynamo with $\mathrm{Pm} =20$, the wavenumber peak for the $U2B$ transfer shifts to lower $b$-shells with time, unlike somewhat fixed $U2B$ wavenumber peak for dynamo with $\mathrm{Pm} =0.2$.  One of the most significant differences between the small-$\mathrm{Pm}$ and large-$\mathrm{Pm}$ dynamos is the nature of the $U2B$ energy transfers.

\begin{figure}[htbp]
\centering
\includegraphics[scale=0.5]{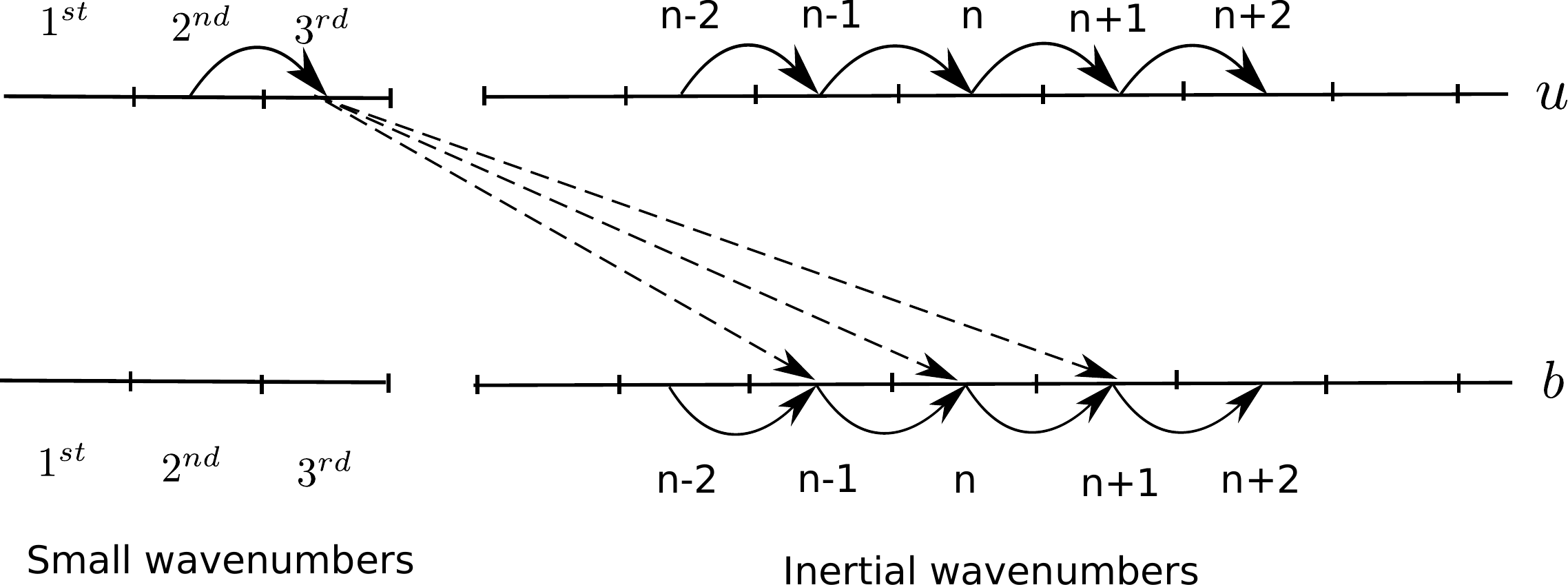} 
\caption{For dynamo with  $\mathrm{Pm} =20$: A schematic diagram of shell-to-shell energy transfers among the $u$-shells and the $b$-shells. The nonlocal transfers are represented by dashed lines.}
\label{fig:ssd_shell_diag}
\end{figure}

A schematic representation of the shell-to-shell energy transfers  for dynamo with $\mathrm{Pm} =20$ is shown in Fig.~\ref{fig:ssd_shell_diag}. It indicates  local and forward $U2U$ and $B2B$ energy transfers, and nonlocal $U2B$ energy transfer.  The most striking difference between the small-Pm and large-$\mathrm{Pm}$ dynamos is the nonlocal $U2B$ energy transfers in the large-$\mathrm{Pm}$ dynamos, and  local $U2B$ energy transfers in the small-Pm dynamos (see Fig.~\ref{fig:lsd_shell_diag}).   

The above difference in the shell-to-shell energy transfer is reflected in the energy fluxes as well.  We observe that in dynamo for $\mathrm{Pm} =0.2$, the magnetic energy growth is dominated by $\Pi^{u<}_{b<}$ energy flux, i.e., by a direct energy transfer from the large-scale velocity field to the large-scale magnetic field.  On the other hand, in dynamo with $\mathrm{Pm} =20$, the flux $\Pi^{u<}_{b>}$ is most dominant, and the kinetic  energy flows from the large-scale velocity field to the small-scale magnetic field.  

Since $\mathrm{Pm}=\nu/\eta$, the magnetic diffusion is stronger for small-Pm dynamo than the large-Pm dynamo.  Therefore,  the magnetic energy at  large-wavenumbers dissipates more strongly for the small-Pm dynamo than the large-Pm dynamo that leads to a spread in the magnetic energy for large Pm dynamos.  Stronger magnetic field at larger wavenumbers yields stronger  $U2B$ energy transfers at large wavenumbers; this effects  more nonlocal  $U2B$ energy transfers for the large-Pm dynamos than the small-Pm dynamos.


In the next section, we perform decaying simulations for $\mathrm{Pm}=20$ and 0.2, and compare the results of decaying and forced simulations.

\section{Decaying simulations}
\label{sec:dec_sim}
Decaying dynamos are encountered when the external forcing is turned off.  We observe such flows in astrophysics, e.g., dynamo action after a huge explosion like supernova, which acts as an energy input at $t=0$.  Decaying dynamos are somewhat generic, and they have relevance to  MHD turbulence as well. 

\begin{figure}[htbp]
\centering
\includegraphics[scale=0.6]{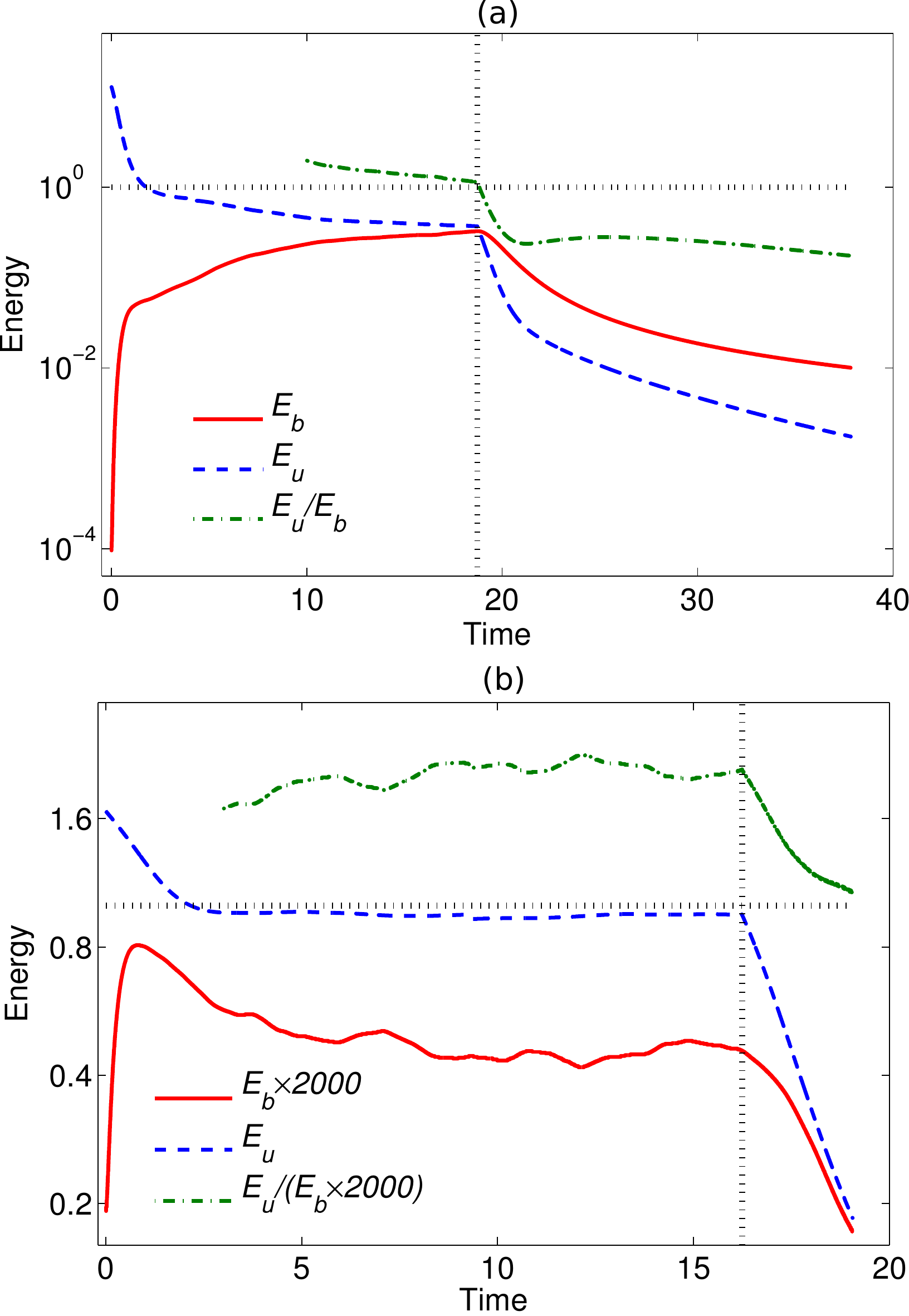}  
\caption{Evolution of the kinetic energy ($E_u$), the magnetic energy ($E_b$), and the ratio $E_u/E_b$ (a) for a dynamo with $\mathrm{Pm} =20$ and (b) for a dynamo with $\mathrm{Pm} =0.2$.  The dynamo for $\mathrm{Pm} =20$ is forced till $t=18.73$ (the dotted vertical line), after which the forcing is turned off. For simulation with $\mathrm{Pm} =0.2$, the forcing is turned off at $t=16.24$.  For  $\mathrm{Pm} =0.2$, $E_b \ll E_u$, hence $E_b$ is multiplied by 2000 to fit in the same plot.}
\label{fig:ssd_lsd_dec_kin_mag}
\end{figure}

First we present the results of our decaying simulation for $\mathrm{Pm}=20$.  We take the output of the forced run at $t=18.73$, and provide it as an input for the decaying run (${\mathbf F} = 0$ in Eq.~(\ref{eq:MHD_vel})).  The initial $E_u/E_b  \approx 1.13$ at $t=18.73$.  The evolution of $E_u$ and $E_b$ with time are exhibited in Fig.~\ref{fig:ssd_lsd_dec_kin_mag}(a) which shows that the kinetic energy decays faster than the magnetic energy, and the $E_u/E_b  \approx 0.17$ at $t_\mathrm{final} = 37.85$. It is interesting to observe that for the decaying dynamo, first $E_u/E_b$ decreases rapidly with time, and then it  tends to flatten out (see the green curve of Fig.~\ref{fig:ssd_lsd_dec_kin_mag}(a)).

\begin{figure}[htbp]
\centering
\includegraphics[scale=1.0]{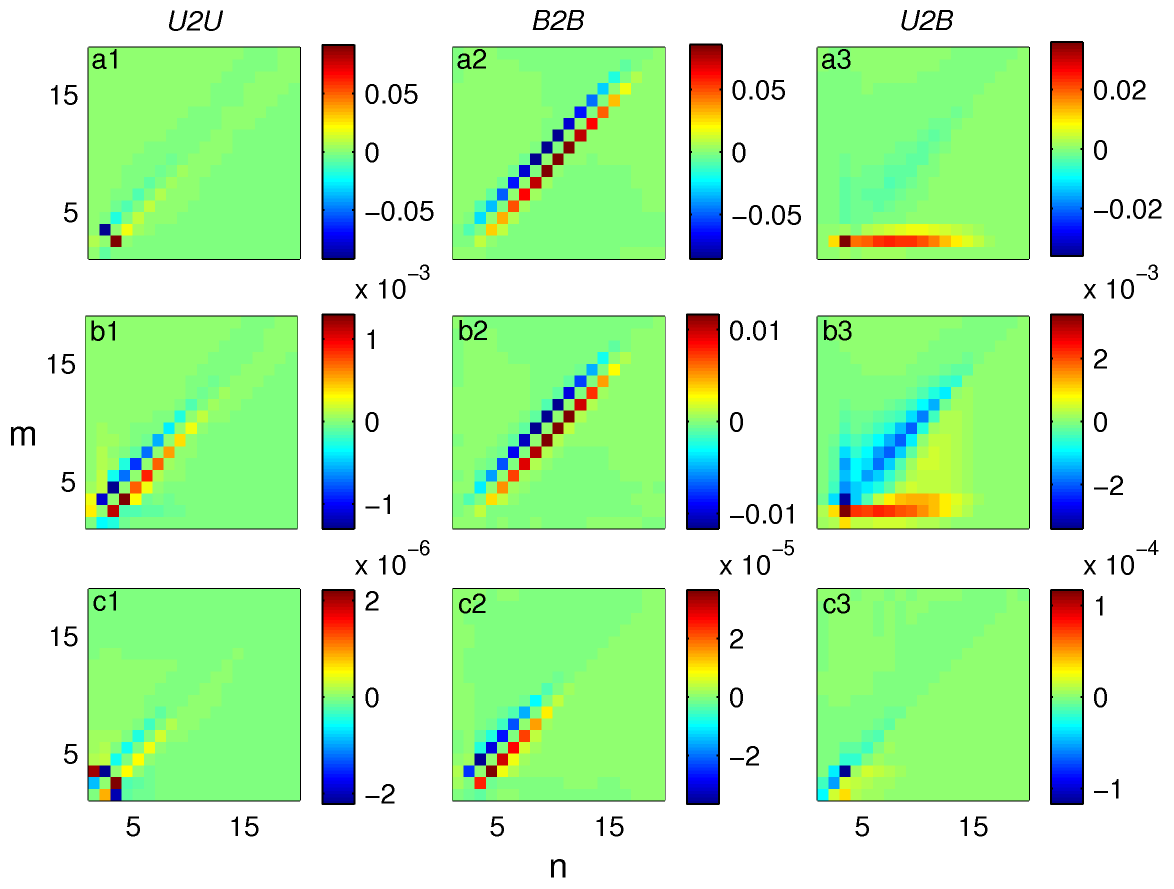}  
\caption{For decaying dynamo with  $\mathrm{Pm} =20$: Shell-to-shell energy transfer rates at $t=18.86$ ((a1)-(a3)),  $t=20.21$ ((b1)-(b3)),  and $t=37.85$ ((c1)-(c3)). The notation is the same as in Fig.~\ref{fig:lsd_shell_uu_bb_ub}. In the initial stages ($t=18.86$), the $U2B$ energy transfer is nonlocal and positive, but at later times ($t=20.21$ and $t=37.85$), the $U2B$ energy transfer becomes local and predominantly negative.}
\label{fig:ssd_dec_shell_uu_bb_ub}
\end{figure}

The shell-to-shell energy transfers for several snapshots during the evolution are shown in Fig.~\ref{fig:ssd_dec_shell_uu_bb_ub}. The $U2U$ and $B2B$ energy transfers are forward and local, similar to what is observed in the corresponding forced simulation. As the simulation progresses, the $U2B$ energy transfer shows some interesting features; the $U2B$ transfer is nonlocal in the early stages of the simulation, as shown in Fig.~\ref{fig:ssd_dec_shell_uu_bb_ub}(a3) (for $t =18.86$), but quickly develops a strong local nature, as shown in Fig.~\ref{fig:ssd_dec_shell_uu_bb_ub}(b3) (for $t =20.21$).  The $U2B$ transfer is fully local in the final stages, e.g., at $t=37.85$.  At $t =20.21$, the ratio of the total local $U2B$ transfers and the total nonlocal $U2B$ transfers  is approximately $4$.  The transition from nonlocal to local energy transfer when forcing is turned off provides interesting clues on the correlations between the velocity and magnetic fields.   We conjecture that the forcing at small wavenumbers induces correlations between the large-scale velocity field and the small- and intermediate-scale magnetic fields; these correlations yield a nonlocal $U2B$ energy transfers.   The above correlations may be too small in decaying MHD, which is the reason for local $U2B$ transfer in decaying dynamo.

The $U2B$ transfer exhibits another interesting phenomenon. In the initial stages of the simulation, the $U2B$ transfer  is positive (i.e., energy transfers from $E_u(k)$ to $E_b(k)$, as shown in Fig.~\ref{fig:ssd_dec_shell_uu_bb_ub}(a3) at $t=18.86$), but at later stages,  $U2B$ transfer changes sign from positive to negative. In the final stages of the simulation, the $U2B$ transfer is predominantly negative (see  Fig.~\ref{fig:ssd_dec_shell_uu_bb_ub}(b3,c3)). Note that $E_u/E_b >1$ in early stages, but   $E_u/E_b < 1$ at later stages.

The asymptotic approach of $E_u/E_b$ in the solar wind as well as in many numerical simulations~\cite{Debliquy:PP2005} with $\mathrm{Pm} = 1$  lies between $0.4-0.6$.  Verma {\it et al.}~\cite{Verma:PP2005}, Debliquy {\it et al.}~\cite{Debliquy:PP2005},  and Verma~\cite{Verma:PR2004} showed  that in kinetically-dominated MHD (large $E_u$), the shell-to-shell energy transfer is preferentially from the velocity field to the magnetic field, and the trend is reversed in magnetically-dominated MHD (large $E_b$).  They attribute the equipartition in turbulent MHD to the above transfers.  We observe similar trends in our decaying simulations, which we claim is the reason for transition from kinetic-energy dominated MHD at $t=18.86$ to magnetic-energy dominated MHD at a later time.  Note however that  in our decaying simulation, the asymptotic $E_u/E_b \approx 0.17$, which is somewhat lower than that observed in the solar wind and in numerical simulations with $\mathrm{Pm=1}$.


\begin{figure}[htbp]
\centering
\includegraphics[scale=0.03]{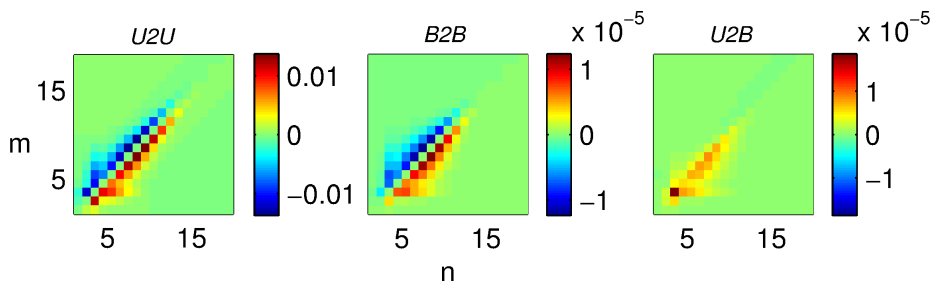} 
\caption{For decaying dynamo with  $\mathrm{Pm} =0.2$: Shell-to-shell energy transfer rates  at $t=19.06$. The notation is the same as in Fig.~\ref{fig:lsd_shell_uu_bb_ub}. The energy transfers are local and forward.}
\label{fig:lsd_dec_shell_uu_bb_ub}
\end{figure} 


\begin{figure}[htbp]
\centering
\includegraphics[scale=0.5]{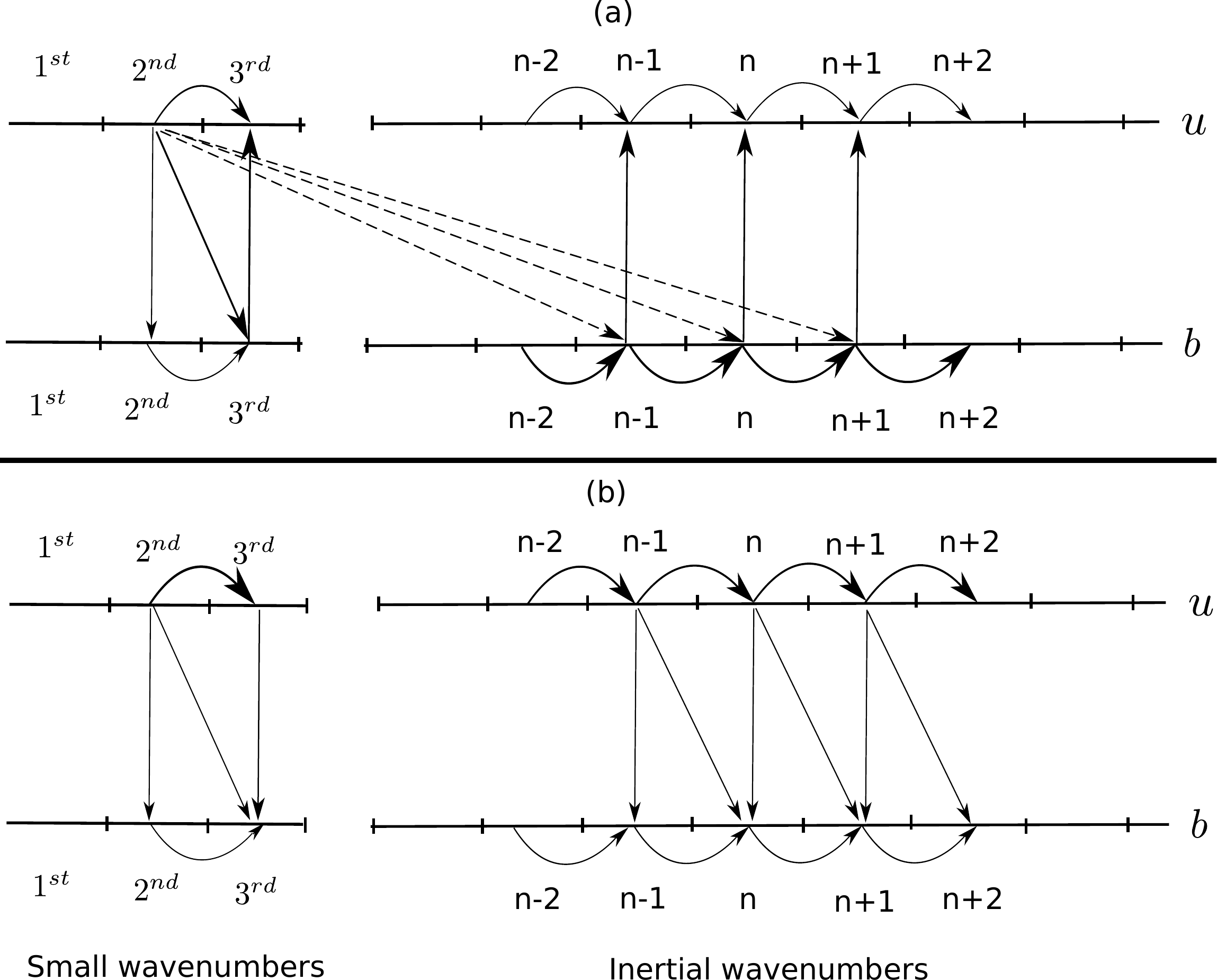}  
\caption{Schematic representation of the shell-to-shell energy transfers among the $u$-shells and the $b$-shells in the decaying dynamo simulation for (a) $\mathrm{Pm} =20$ and (b) $\mathrm{Pm} =0.2$.}
\label{fig:ssd_lsd_dec_shell_diag}
\end{figure}

We also perform the decaying dynamo simulation for $\mathrm{Pm} =0.2$.  For the initial condition, we use the corresponding forced simulation state at $t= 16.24$.  The time-evolution of its $E_u$ and $E_b$  are exhibited in Fig.~\ref{fig:ssd_lsd_dec_kin_mag}(b), according to which the kinetic energy decays somewhat faster than the magnetic energy, with the ratio $E_u/E_b$ varying from  approximately $4000$ at $t=16.24$ to $2100$ at $t=19.06$.  The convergence of the decaying dynamo with $\mathrm{Pm} =0.2$ is very slow, hence the asymptotic state is not reported in this paper.

The nature of energy transfers in the decaying simulations for $\mathrm{Pm} =20$ and $0.2$ are quite different.   The shell-to-shell energy transfers during the decay of $\mathrm{Pm} = 0.2$   dynamo  are shown in Fig.~\ref{fig:lsd_dec_shell_uu_bb_ub}. The $U2U$ and $B2B$ energy transfers are forward and local whereas the $U2B$ transfers are forward and local with a small nonlocal component, same as observed in the forced simulation for $\mathrm{Pm} =0.2$. The energy transfer from kinetic to magnetic for $\mathrm{Pm}=0.2$ is due to the dominance of the kinetic energy over the magnetic energy, in contrast to that observed for $\mathrm{Pm}=20$.  Thus, in the forced and decaying simulations for $\mathrm{Pm} =0.2$, we observe similar kinds of energy transfers in contrast to the simulation for $\mathrm{Pm} =20$, for which the $U2B$ energy transfers are different in the decaying and forced simulations. 

The schematic diagrams of the shell-to-shell energy transfers in the decaying simulations for $\mathrm{Pm} =20$ and $0.2$ are shown in Fig.~\ref{fig:ssd_lsd_dec_shell_diag}(a) and Fig.~\ref{fig:ssd_lsd_dec_shell_diag}(b), respectively. It shows local and forward $U2U$ and $B2B$ energy transfers in the decaying simulation for $\mathrm{Pm} =20$ (shown in Fig.~\ref{fig:ssd_lsd_dec_shell_diag}(a)). However, the $U2B$ transfers consists of nonlocal transfers from $u$-shell to $b$-shells, and local transfers from $b$-shells to $u$-shells. For $\mathrm{Pm} =0.2$, however, the energy transfers in the decaying simulation, shown in Fig.~\ref{fig:ssd_lsd_dec_shell_diag}(b), are similar to those of the corresponding forced simulation,  shown in Fig.~\ref{fig:lsd_shell_diag}.

\section{Discussions and conclusions}
\label{sec:conclude}

In this paper, we study the growth of the magnetic energy in dynamo for $\mathrm{Pm} <1$ using numerical simulations.  We perform direct numerical simulation for $\mathrm{Pm}=0.2$ on $1024^3$ grid; the magnetic Reynolds number for the steady state is 181, which is near the dynamo transition regime.  We study energy fluxes and shell-to-shell energy transfers using the numerical data.  In addition, we also perform a lower resolution dynamo simulation for $\mathrm{Pm} =0.2$ (but with larger $\nu$ and $\eta$) on $256^3$ grid, with a lower $\mathrm{Rm}$. We compare the energy transfers in the forced MHD simulation for $\mathrm{Pm} =0.2$ with that of $\mathrm{Pm} =20$. We  also emphasize the differences between energy transfers in forced and decaying simulations for the two types of dynamos. The summary of findings are as follows:
\begin{enumerate}

\item In the forced dynamo simulation for $\mathrm{Pm}=0.2$ on $1024^3$ grid, the kinetic and magnetic energies saturate quickly in several eddy turnover time, with the magnetic energy $4000$ times smaller than the kinetic energy.  The deviation from the equipartition is because the dynamo is near the dynamo onset which allows $E_b/E_u$ very different from unity.

\item In dynamo for $\mathrm{Pm} =0.2$, the magnetic field growth takes place due to predominantly local energy transfers from the large-scale velocity field to the large-scale magnetic field.

\item  In the forced dynamo simulation for $\mathrm{Pm} =0.2$ on $256^3$ grid, the kinetic and magnetic energies saturates to the asymptotic state of $E_b/E_u \approx 1/5$.

\item The nature of energy transfers during the growth of dynamo for $\mathrm{Pm} =0.2$ is approximately the same for both $1024^3$ and $256^3$ resolutions. Note that $\nu$ and $\eta$ for the higher-resolution grids is lower than those for the lower-resolution ones.

\item  In the forced simulation for $\mathrm{Pm} =20$, the growth of magnetic energy takes place due to the nonlocal energy transfers from the large-scale velocity field to the small-scale magnetic field. In contrast, for the decaying dynamo with $\mathrm{Pm} =20$, the $U2B$ energy transfer is predominantly local, and its direction is from  $B$ to $U$.

\item  In the decaying dynamo simulation for $\mathrm{Pm} =0.2$, the $U2U$, $B2B$, and $U2B$ energy transfers are similar to that of the corresponding forced simulation.  
\end{enumerate}

Here we provide qualitative explanation for the change in behaviour in the energy transfers for the three cases: small Pm, large Pm, and Pm=1. A common feature for the three cases is that the velocity-to-velocity ($U2U$) and magnetic-to-magnetic ($B2B$) energy transfers are forward and local.   This has been reported earlier, specially for $\mathrm{Pm}=1$~\cite{Dar:PD2001,Carati:JT2006,Verma:PR2004}. In this paper we focus on small-Pm and large-Pm dynamos.  

The velocity-to-magnetic ($U2B$) energy transfer however shows different behaviour for the three cases. Here we contrast these cases.  During the early phase of the dynamo growth, the velocity-to-magnetic ($U2B$) energy transfers are forward and local (see Fig.~7).  This is due to the nonlinear term ${\bf b}\cdot \nabla {\bf u}$ that generates larger wavenumber ($k$) modes by nonlinearity.  The dissipation of magnetic energy is very weak since the large $k$ modes are yet to be populated. The $U2B$ energy transfer cascades locally from smaller to larger wavenumbers.

When significant energy has reached the large $k$'s,  dissipation  of magnetic energy commences.  The magnetic diffusion is stronger for small-Pm dynamo than for large-Pm dynamo, hence the magnetic energy at  large-wavenumbers dissipates more strongly for the small-Pm dynamo than the large-Pm dynamo.  Therefore, the magnetic energy at large-$k$ is suppressed more intensely for the small-Pm case than that for large-Pm case.  This is why the magnetic energy is more spread out in the Fourier space for dynamos with large-Pm than those for small-Pm.  

The $U2B$ energy transfer is strong when the magnetic field  is significant.  Therefore, for the large-Pm dynamos, due to the aforementioned scatter of the magnetic energy at large-$k$, the  $U2B$ energy transfer is more nonlocal than that in small-Pm dynamo.  

The $U2B$ energy transfers in MHD turbulence appears to have a significant nonlocality.  This feature has been reported earlier by Carati {\it et al.}~\cite{Carati:JT2006} for Pm=1 dynamo as well. This feature may be due to certain correlations among the velocity, magnetic, and force fields. The above conjecture is strengthened by our decaying dynamo runs where $U2B$ energy transfers become local in the absence of the external forcing.   These topics are beyond the scope of the present paper; these are part of our future analysis.

In conclusion, the energy transfer studies provide valuable insights into the dynamo mechanism for forced as well as decaying MHD simulations.


\section*{Acknowledgements}
We thank Emmanuel Dormy and Binod Sreenivasan for valuable discussions and comments. We are grateful to Madhusudhanan Srinivasan and Shuaib Arshad for helping us with visualization plots. For computer time, this research used the resources of the Supercomputing Laboratory at King Abdullah University of Science and Technology (KAUST) in Thuwal, Saudi Arabia. This work was supported by a research grant SERB/F/3279/2013-14 from Science and Engineering Research Board, India.  RS was supported through baseline funding at KAUST.



\end{document}